\documentclass[aps,prd,onecolumn,notitlepage,nofootinbib,showpacs,preprintnumbers,superscriptaddress,groupedaddress]{revtex4-1}
\usepackage[T1]{fontenc}
\usepackage[utf8]{inputenc}
\usepackage{bm}
\usepackage{amsmath}
\usepackage{amssymb}
\usepackage{esint}
\usepackage{color}
\usepackage{array}
\usepackage{float}
\usepackage{graphicx}
\PassOptionsToPackage{normalem}{ulem}
\usepackage{ulem}

\usepackage{hyperref}

\begin{document}
	
	\title{Higher derivative scalar-tensor monomials and their classification}
	
	\author{Xian Gao}%
	\email[Email: ]{gaoxian@mail.sysu.edu.cn}
	\affiliation{%
		School of Physics and Astronomy, Sun Yat-sen University, Guangzhou 510275, China}
	
	\date{March 26, 2020}
	
	\begin{abstract}
		We make a full classification of scalar monomials built of the Riemann curvature tensor up to the quadratic order and of the covariant derivatives of the scalar field up to the third order. From the point of view of the effective field theory, the third or even higher order covariant derivatives of the scalar field are of the same order as the higher curvature terms, and thus should be taken into account. Moreover, higher curvature terms and higher order derivatives of the scalar field are complementary to each other, of which novel ghost-free combinations may exist. We make a systematic classification of all the possible monomials, according to the numbers of Riemann tensor and higher derivatives of the scalar field in each monomial. Complete basis of monomials at each order are derived, of which linear combinations may yield novel ghost-free Lagrangians. We also develop a diagrammatic representation of the monomials, which may help to simplify the analysis.
	\end{abstract}
	
	\maketitle


\section{Introduction}

Scalar-tensor theory is extensively studied in the past few decades as one of the main theories of modification of gravity.
In particular, much effort has been made to introduce higher derivatives of the scalar field without the Ostrogradsky ghost(s) \cite{Woodard:2015zca}. 
The representative achievements are the Horndeski theory \cite{Horndeski:1974wa,Nicolis:2008in,Deffayet:2011gz,Kobayashi:2011nu} as well as the  degenerate higher-order theory  \cite{Gleyzes:2014dya,Gleyzes:2014qga,Langlois:2015cwa,Motohashi:2016ftl} (see Refs. \cite{Langlois:2018dxi,Kobayashi:2019hrl} for reviews).  

However, the previous studies mostly stopped at the second order in the derivative of the scalar field\footnote{Non-polynomial derivative terms that are infinite order in derivatives have also been studied, see (e.g.,) \cite{Buoninfante:2018lnh} and references therein.}. This is partly because the second derivatives of the scalar field provide a ``playground'' of higher derivative scalar-tensor theory that is sufficiently nontrivial but not too exhausting to be studied. 
Although focusing on the second derivative is consistent by itself, there are at least two motivations to go beyond, as we shall explain below.

First, from the point of view of the effective field theory, operators of the same order in derivative are of the same importance and should be treated in the same footing.
As an example, let us recall that $\mathcal{L}_{4}^{\text{H}}$ of the Horndeski theory with $G_{4} = X^2$ takes the form
	\begin{equation}
	\mathcal{L}_{4}^{\text{H}} = X^{2}\, R + 2X\left(\square\phi\right)^{2}-2X\left(\nabla_{a}\nabla_{b}\phi\right)^{2},  \label{LH4}
	\end{equation}
with $X= - \frac{1}{2} \nabla_{a}\phi\nabla^{a}\phi$. Since the Riemann tensor arises from the commutator of two covariant derivatives, we may think that $R \sim \mathcal{O}(\nabla^2)$. 
As a result, all the three terms in the above are of $\mathcal{O}(\nabla^6)$. This partly explains why these three terms arise together in the Horndeski theory.
On the other hand, there are other terms that are of $\mathcal{O}(\nabla^6)$ as well.
Schematically, one example is
	\[
		\sim X \nabla_{a}\phi \nabla^{a} \square \phi,
	\]
in which the third derivative of the scalar field arises.
Of course, for this particular case (i.e., $\mathcal{L}_{4}^{\text{H}}$), this term is trivial since it can be reduced to the Horndeski form by integrations by parts.
Nevertheless, things become less trivial if we consider $\mathcal{L}_{5}^{\text{H}}$ with
	\[
	\mathcal{L}_{5}^{\text{H}}=X^{2}\, G^{ab}\nabla_{a}\nabla_{b}\phi-\frac{1}{3}X\left(\square\phi\right)^{3}+ X\square\phi\left(\nabla_{a}\nabla_{b}\phi\right)^{2}-\frac{2}{3}X\left(\nabla_{a}\nabla_{b}\phi\right)^{3}, 
	\]
where $G_{ab}$ is the Einstein tensor. All the terms in the above are $\sim \mathcal{O}(\nabla^{8})$. 
There is another type of terms, schematically
	\[
		\sim \nabla\phi\nabla\phi\nabla\phi \nabla \nabla \phi \nabla \nabla \nabla \phi,
	\]
which are also $\sim \mathcal{O}(\nabla^{8})$ and thus should be consider in the same footing.
In particular, this type of terms cannot be fully reduced by integrations by parts.

The second motivation comes from the study of higher derivative scalar-tensor theories, and in particular, higher curvature gravity theories that are ghostfree in the so-called unitary gauge, in which the scalar field is chosen to be spatially uniform.
Perhaps in this sense the simplest example is the Chern-Simons gravity \cite{Lue:1998mq,Jackiw:2003pm}, which suffers from ghosts but is ghostfree in the unitary gauge.
Another interesting example was introduced in \cite{Deruelle:2012xv}, which is quadratic in the Weyl tensor and is ghostfree in the unitary gauge.
Some exotic parity-violating scalar-tensor theories that are healthy in the unitary gauge were identified in \cite{Crisostomi:2017ugk}, in which terms quadratic in the curvature tensor also arise.
Although there are still debates on their behaviour in a general background \cite{DeFelice:2018mkq}, such kind of theories provide us an even broader framework of higher derivative scalar-tensor theories that have many applications.
Refs. \cite{Deruelle:2012xv,Crisostomi:2017ugk} focus on the direct couplings between the curvature and derivatives of the scalar field up to the second order.
According to the same reason in the above, terms built of derivatives of the scalar field higher than the second order should also be considered.
For example, terms that are quadratic in the Riemann tensor were introduced in \cite{Deruelle:2012xv,Crisostomi:2017ugk}, which take the schematic form
	\[
		\sim  \underbrace{\nabla\phi \cdots \nabla\phi}_{n} R\,R, 
	\]
where $R$ is a shorthand for the Riemann tensor, and there are $n$ first derivatives of the scalar field. This type of terms is of $\sim \mathcal{O}(\nabla^{n+4})$. While terms in the form
	\[
		\sim \underbrace{\nabla\phi \cdots \nabla\phi}_{n-2} \nabla\nabla\nabla \phi \nabla\nabla\nabla \phi
	\]
have the same order and thus should be considered as well.

From another point of view, terms with derivatives of the scalar field higher than the second order can be treated as being ``complimentary'' to the higher curvature terms.
This is similar to case of Horndeski theory, in which the Lagrangian can be split into the ``curvature sector'' and the ``scalar field sector'', and one is complimentary to the other\footnote{This is similar to the ``covariantization'' procedure in \cite{Deffayet:2009wt,Deffayet:2011gz}.}.
For example, in $\mathcal{L}_{4}^{\text{H}}$, the ``curvature sector'' is $ X^{2}R$ and the ``scalar field sector'' is $2X\left(\square\phi\right)^{2}-2X\left(\nabla_{a}\nabla_{b}\phi\right)^{2}$. Neither the curvature nor the scalar field sector can be tuned to be ghostfree individually. Only their linear combination can yield a ghostfree covariant Lagrangian.
Then it is natural to ask what the complementary terms of (e.g.) quadratic curvature terms are, and whether their combinations can yield new ghostfree theory?

In order to answer the above questions, a systematic investigation of more general higher derivatives of the scalar field and their couplings with the curvature is required.
As a first step, this paper is devoted to the classification of monomials built of derivatives of the scalar field up to the third order as well as their couplings with the curvature tensor.

The paper is organized as following.
In Sec. \ref{sec:form}, we setup our formalism by explaining the necessity of including derivatives of the scalar field higher than the second order. Then we make the classification of all the monomials according to their derivatives.
We then construct all the monomials for $d=1,2,3,4$ in Sec. \ref{sec:d1}, \ref{sec:d2}, \ref{sec:d3} and \ref{sec:d4}, respectively.
Sec. \ref{sec:con} concludes.

\section{The formalism} \label{sec:form}

A general Lagrangian that is built of a single scalar field and the Riemann tensor as well as their covariant derivatives takes the form
	\begin{equation}
		\mathcal{L}\left(g^{ab},\varepsilon_{abcd};R_{abcd},\phi,\nabla_{a}\right),
	\end{equation}
where the complete antisymmetric Levi-Civita tensor is
	\begin{equation}
	\varepsilon_{abcd} = \sqrt{-g} \, \epsilon_{abcd},
	\end{equation}
with $\epsilon_{0123}=1$.
The form of the Lagrangian can be quite arbitrary in general.
In this paper we concentrate on the case that the Lagrangian is a polynomial, in which the monomials are scalar invariants that are built of the the Riemann curvature tensor, the scalar field and their covariant derivatives, with possible coefficients of the form $X^{n}$, where
	\begin{eqnarray}
	X= - \frac{1}{2} \nabla_{a}\phi\nabla^{a}\phi,
	\end{eqnarray}
is the canonical kinetic term of the scalar field.

Precisely, each monomial takes the general structure\footnote{We use the word ``monomial'' since it is a scalar built of the products of several tensors with all the indices being contracted. Of course, in the sense of tensor components, each monomial is in fact a summation of many terms.}
	\begin{equation}
	\underbrace{\cdots R\cdots}_{c_{0}}\underbrace{\cdots\nabla R\cdots}_{c_{1}}\underbrace{\cdots\nabla\nabla R\cdots}_{c_{2}}\cdots\underbrace{\cdots\nabla\phi\cdots}_{d_{1}}\underbrace{\cdots\nabla\nabla\phi\cdots}_{d_{2}}\underbrace{\cdots\nabla\nabla\nabla\phi\cdots}_{d_{3}}\underbrace{\cdots\nabla\nabla\nabla\nabla\phi\cdots}_{d_{4}}\cdots,
	\end{equation}
where ``$\cdots$'' denotes multiple Riemann curvature tensor (we schematically denote as $R$), the scalar field and their covariant derivatives.
All the indices are contracted by the metric $g^{ab}$ and/or the complete antisymmetric tensor $\varepsilon^{abcd}$. 
Thus, we may assign each monomial a set of integers
	\begin{eqnarray}
		\left(c_{0},c_{1},c_{2},\cdots;d_{1},d_{2},d_{3},d_{4},\cdots\right),
	\end{eqnarray}
where 
\begin{itemize}
	\item $c_{0},c_{1},c_{2},\cdots$ are numbers of Riemann curvature tensor and its first, second derivatives, etc.,
	\item $d_{1},d_{2},d_{3},d_{4},\cdots$ are numbers the first, the second, the third and the fourth covariant derivatives of $\phi$, etc..
\end{itemize}
We assume all the $c_{n}$'s and $d_{n}$'s are non-negative except $d_{1}$. This is simply because we allow the monomials to be divided by some powers of $X$, which will be convenient in the formalism we shall develop.
As an example, the term
	\[
	\sim X\,R^2\,\nabla_{a}R\,\nabla^{a}\square\phi\left(\square\phi\right)^{2}
	\]
corresponds to
	\[
	\left(2,1,0,\cdots;2,2,1,0,\cdots\right).
	\]

We shall classify various monomials according to the order of derivatives as well as the partition of derivatives.
First, the total number of derivatives of a given monomial with $(c_{0},c_{1},c_{2},\cdots;d_{1},d_{2},d_{3},d_{4},\cdots)$ is
	\[
		D\equiv2c_{0}+3c_{1}+4c_{2}+\cdots+d_{1}+2d_{2}+3d_{3}+4d_{4}+\cdots,
	\]
that is
	\begin{equation}
		D \equiv \sum_{n=0}\left[\left(n+2\right)c_{n}+n\,d_{n}\right], \label{D_def}
	\end{equation}
which is justified by the fact that the Riemann tensor is the commutator of two covariant derivatives.
That is, we will treat each Riemann tensor as order 2 in covariant derivatives, or schematically, $R\sim \mathcal{O}(\nabla^2)$.
Thus a monomial corresponds to $(c_{0},c_{1},c_{2},\cdots;d_{1},d_{2},d_{3},d_{4},\cdots)$ is of $\sim\mathcal{O}(\nabla^D)$ with $D$ given in (\ref{D_def}).

Second, one of the main points of view taken in this work is that different monomials with the same $D$ should be treated as ``of the same order''.
This is also important when one tries to build ghostfree combinations (polynomials) out of various scalar-tensor monomials.
In fact, only polynomials that are linear combinations of monomials with the same $D$ can be tuned to be possibly degenerate and thus ghostfree.

As a simple and illustrative example, let us recall that $\mathcal{L}_{4}^{\text{H}}$ of the Horndeski theory with $G_{4} = X^2$ is given in (\ref{LH4}).
The first term of $\mathcal{L}_{4}^{\text{H}}$ corresponds to $(c_0,c_1,\cdots;d_1,d_2,\cdots) = (1,0,\cdots;4,0,\cdots)$ and the last two terms correspond to $(c_0,c_1,\cdots;d_1,d_2,\cdots) = (0,0,\cdots;2,2,\cdots)$. They all correspond to $D=6$ and thus arise together with fine tuned coefficients to make a ghostfree polynomial.

This counting, although is applicable by itself, does not fully capture the crucial structure of the monomials as well as their combinations, which is encoded in the higher derivatives of the scalar field.
In fact, the number of the first derivatives $d_{1}$ is not crucial as the first derivatives will not affect the degeneracy structure of the theory. 
Let us take $\mathcal{L}_{4}^{\text{H}}$ in (\ref{LH4}) as an example again.
Alternatively, we may think this specific Lagrangian in the form
	\[
		\mathcal{L}_{4}^{\text{H}}=X^{2}\times\left\{ R+\frac{2}{X}\left[\left(\square\phi\right)^{2}-\left(\nabla_{a}\nabla_{b}\phi\right)^{2}\right]\right\},
	\]
where in the curly bracket, all the monomials are recast in a form such that the dimensions of the scalar field are completely cancelled\footnote{That is, each monomial is invariant under $\phi \rightarrow \lambda \phi$ with constant $\lambda$.}.
The trick is simply to divide each derivative term of the scalar field by the factor $\sim \nabla\phi$, i.e., schematically $\nabla\nabla\phi \rightarrow \frac{1}{\nabla\phi}\nabla\nabla\phi$, $\nabla\nabla\nabla\phi \rightarrow \frac{1}{\nabla\phi}\nabla\nabla\nabla\phi$, etc..
As a result, instead of using $D$ directly, is convenient to define another number
	\[
		d = D-(d_1+d_2+d_3+d_4+\cdots) ,
	\]
that is
	\begin{equation}
		d \equiv \sum_{n=0}\left[\left(n+2\right)c_{n}+\left(n+1\right)\,d_{n+2}\right]. \label{d_def}
	\end{equation}
Since $d_{1}$ completely drops out in $d$, from now on we may suppress $d_{1}$ and write $(c_{0},c_{1},\cdots;d_{2},d_{3},\cdots)$.

Another motivation of using $d$ instead of $D$ comes from the relation between the covariant scalar-tensor Lagrangian and the corresponding Lagrangian in the so-called unitary gauge.
When fixing the unitary gauge, the generally covariant scalar-tensor theory can be written in terms of spatially covariant gravity theories, in which the basic building blocks are spatially covariant tensors, such as the spatial metric $H_{ab}$, extrinsic curvature $K_{ab}$ and intrinsic curvature ${}^{3}\!R_{ab}$ as well as their spatial and temporal derivatives.
Spatially covariant theories of gravity have been proposed and studied previously with different motivations and forms, such as in the effective field theory of inflation \cite{Creminelli:2006xe,Cheung:2007st}, in the Ho\v{r}ava gravity \cite{Horava:2009uw,Blas:2009qj}, etc.
Spatially covariant gravity theories with at most three degrees of freedom were extensively studied in \cite{Gao:2014soa,Gao:2014fra,Fujita:2015ymn,Gao:2018znj,Gao:2019lpz,Gao:2018izs,Gao:2019liu,Gao:2019twq}.

One important observation is that when transferring to the unitary gauge, the number of derivatives (in the resulting spatially covariant gravity terms) is reduced by one, comparing with the original covariant derivatives of the scalar field.
In fact, the correspondences of the first and the second order derivatives of the scalar field in the unitary gauge are given by\footnote{Here and in what follows we have not chosen the hypersurface-adapted coordinates, and all the quantities are still 4-dimensional tensors with spacetime indices.}
	\begin{eqnarray}
	\nabla_{a}\phi & \rightarrow & -\frac{1}{N}u_{a},\label{dec_dp1}\\
	\nabla_{a}\nabla_{b}\phi & \rightarrow & u_{a}u_{b}A-2u_{(a}B_{b)}+\Delta_{ab},\label{dec_dp2}
	\end{eqnarray}
with
	\begin{eqnarray}
	A & \equiv & -\frac{1}{N}\pounds_{\bm{u}}\ln N,\\
	B_{a} & \equiv & -\frac{1}{N}a_{a},\\
	\Delta_{ab} & \equiv & -\frac{1}{N}K_{ab},
	\end{eqnarray}
where $H_{ab}=g_{ab}+u_{a}u_{b}$, $a_{a}=\pounds_{\bm{u}}u_{a}$ and $K_{ab}=\frac{1}{2} \pounds_{\bm{u}}H_{ab}$ are the induced metric, the acceleration and the extrinsic curvature, respectively.
For the third derivative of the scalar field, we find \cite{Gao:2015xwa}
	\begin{equation}
	\nabla_{c}\nabla_{a}\nabla_{b}\phi\rightarrow-u_{c}u_{a}u_{b}\,U+3u_{(c}u_{a}V_{b)}-u_{c}X_{ab}-2Y_{c(a}u_{b)}+Z_{cab},\label{dec_dp3}
	\end{equation}
with
	\begin{eqnarray}
	U & = & \frac{1}{N}\left[\left(\pounds_{\bm{u}}\ln N\right)^{2}-\pounds_{\bm{u}}^{2}\ln N+2a^{d}a_{d}\right],\\
	V_{b} & = & -\frac{1}{N}\left(-2a_{b}\pounds_{\bm{u}}\ln N+\pounds_{\bm{u}}a_{b}-2a_{d}K_{b}^{d}\right),\\
	X_{ab} & = & \frac{1}{N}\left(K_{ab}\pounds_{\bm{u}}\ln N+2a_{a}a_{b}+2K_{a}^{d}K_{bd}-\pounds_{\bm{u}}K_{ab}^{(u)}\right),\\
	Y_{cb} & = & \frac{1}{N}\left[K_{cb}\pounds_{\bm{u}}\ln N-\mathrm{D}_{c}a_{b}+a_{c}a_{b}+K_{c}^{d}K_{db}\right],\\
	Z_{cab} & = & \frac{1}{N}\left(-\mathrm{D}_{c}K_{ab}+3a_{(c}K_{ab)}\right).
	\end{eqnarray}
In the above, $\pounds_{\bm{u}}$ is the Lie derivative with respect to $u_{a}$ and $\mathrm{D}_{a}$ is the projected derivative with respect to the induced metric.
It is thus transparent that $\nabla_{a}\phi$, $\nabla_{a}\nabla_{b}\phi$ and $\nabla_{c}\nabla_{a}\nabla_{b}\phi$ correspond to $\sim \mathcal{O}(\nabla^{0})$, $\sim \mathcal{O}(\nabla^{1})$ and $\sim \mathcal{O}(\nabla^{2})$ in the unitary gauge, respectively.
Therefore the number $d$ can also be viewed as the total number of derivatives of the corresponding terms in the unitary gauge.
Using $d$ thus makes the connection between covariant scalar-tensor invariants and spatially covariant gravity transparent \cite{Gao:toappear}.

In this work, we consider monomials with $d\leq 4$. It is thus eligible to consider $c_{n}$'s up to $c_{2}$ and $d_{n}$'s up to $d_{4}$.
We then classify all the possible categories of a given $d$ with different $c_{n}$'s and $d_{n}$'s. The results are summarized in Table \ref{tab:clsST}.
	\begin{table}[H]
		\begin{centering}
			\begin{tabular}{|c|>{\centering}p{3cm}|>{\centering}p{3cm}|}
				\hline 
				& Irreducible & Reducible\tabularnewline
				\hline 
				$d$ & $\left(c_{0};d_{2},d_{3}\right)$ & $\left(c_{0},c_{1},c_{2};d_{2},d_{3},d_{4}\right)$\tabularnewline
				\hline 
				1 & $\left(0;1,0\right)$ & -\tabularnewline
				\hline 
				2 & $\left(0;2,0\right)$
				
				$\left(1;0,0\right)$ & $\left(0,0,0;0,1,0\right)$\tabularnewline
				\hline 
				3 & $\left(0;3,0\right)$
				
				$\left(0;1,1\right)$
				
				$\left(1;1,0\right)$ & $\left(0,0,0;0,0,1\right)$
				
				$\left(0,1,0;0,0,0\right)$\tabularnewline
				\hline 
				4 & $\left(0;4,0\right)$
				
				$\left(0;2,1\right)$
				
				$\left(0;0,2\right)$
				
				$\left(1;2,0\right)$
				
				$\left(2;0,0\right)$
				
				$\left(1;0,1\right)$ & $\left(0,0,0;1,0,1\right)$
				
				$\left(0,1,0;1,0,0\right)$
				
				$\left(0,0,1;0,0,0\right)$\tabularnewline
				\hline 
			\end{tabular}
			\par\end{centering}
		\caption{Classification of scalar-tensor monomials up to $d=4$.}		
		\label{tab:clsST}
	\end{table}
Comments are in order.
	\begin{enumerate}
		\item We split all the categories into to cases, which we dub as the irreducible and reducible cases, respectively.
		Monomials belong to the reducible cases can be reduced to (the linear combinations of) monomials belonging to the irreducible cases.
		In the right column of Table \ref{tab:clsST}, there are 6 reducible categories. For examples, the category $\left(0,0,0;0,1,0\right)$, which corresponds to monomials that are of the schematic form
			\[
			\nabla\phi \cdots \nabla\phi \nabla\nabla\nabla\phi
			\]
		can always be reduced 
		\[
		\nabla\phi \cdots \nabla\phi \nabla\nabla\nabla\phi \simeq \nabla\phi \cdots \nabla\phi \nabla\nabla\phi \nabla\nabla\phi,
		\]
		up to total derivatives.
		In this sense we refer to the category $\left(0,0,0;0,1,0\right)$ as being reducible and schematically write
		\[
		(0,0,0;0,1,0) \simeq (0,0,0;2,0,0).	
		\]
		It is thus easy to verify that all the 6 categories in the right column in Table \ref{tab:clsST} can be reduced by integrations by parts.
		
		\item  We notice that all the irreducible categories in Table \ref{tab:clsST} have $c_{1}=c_{2} = d_{4}=0$. For the sake of briefness, we therefore suppress $c_{1},c_{2} ,d_{4}$ in their notation and thus use 3 integers to denote
		\begin{equation}
		(c_0;d_2,d_3) \equiv (c_0,0,0;d_2,d_3,0)
		\end{equation}
		in the middle column of Table \ref{tab:clsST} and in the rest part of this paper.
		Thanks to this fact, up to $d=4$ we do not need to consider the fourth order derivatives of the scalar field, neither the derivatives of the Riemann tensor.
		
		\item The necessity of including the third order derivatives of the scalar field $\nabla\nabla\nabla\phi$ becomes transparent according to the Table \ref{tab:clsST}. On one hand, monomials with $\nabla\nabla\nabla\phi$ can be of the same order as monomials with only up to the second order derivatives. For example, $(0;1,1)\sim(0;3,0)$ and $(0;2,1)\sim(0;0,2)\sim(0;4,0)$. On the other hand, as has been discussed in the Introduction, monomials built of higher derivatives of the scalar field can be viewed as the complementary terms of monomials built of the Riemann tensor with direct couplings with the scalar field. For example, $\mathcal{L}_{4}^{\mathrm{H}}$ shows $(0;2,0) \sim(1;0,0)$ and $\mathcal{L}_{5}^{\mathrm{H}}$ shows $(0;3,0)\sim (1;2,0)$. While the fact that $(0;1,1)\sim(1;1,0)$ and $(0;2,1)\sim(0;0,2)\sim(1;2,0) \sim(2;0,0)\sim(1;0,1)$ implies that there might be more general combinations of higher derivative terms of the scalar field and higher curvature terms that can be ghostfree. In particular, monomials that are quadratic in the curvature tensor were considered in  \cite{Deruelle:2012xv,Crisostomi:2017ugk}, our classification indicates the terms built of higher derivatives of the scalar field may act as  complementary terms such that the novel ghostfree combinations may arise. 
	\end{enumerate}

In the rest part of this paper, we construct all the monomials for each category in the irreducible case according to Table \ref{tab:clsST}.

\section{$d=1$} \label{sec:d1}

The case of $d=1$ is simple and we will use it to illustrate our strategy and formalism.

For $d=1$, according to Table \ref{tab:clsST}, there is only one irreducible category $(0;1,0)$, which corresponds to monomials in which the second order derivatives enter linearly. There are two  monomials, which we denote to be
	\begin{eqnarray}
	\bm{E}_{1}^{\left(0;1,0\right)} & \equiv & \frac{1}{\sigma}\square\phi, \label{E010_1}\\
	\bm{E}_{2}^{\left(0;1,0\right)} & \equiv & \frac{1}{\sigma^{3}}\nabla_{a}\phi\nabla_{b}\phi\nabla^{a}\nabla^{b}\phi, \label{E010_2}
	\end{eqnarray}
where in what follows we denote
	\begin{equation}
	\sigma = \sqrt{2X}
	\end{equation}
for short.

There are several comments we would like to make.
	\begin{enumerate}
		\item Here and throughout this paper, we use the notation $\bm{E}^{(c_0;d_2,d_3)}_{n}$ to denote monomials built of $c_0$ Riemann curvature tensor, $d_2$ second order derivatives and $d_{3}$ third order derivatives of the scalar field. Again, all the indices are contracted by the metric and/or the Levi-Civita tensor.
		\item We deliberately divided the monomials by powers of $\sigma$ such that the resulting monomials are dimensionless with respect to $\phi$. This is especially convenient when one tries to construct ghostfree combinations of several monomials, in which the coefficient are purely numerical constant. 
		\item Both monomials $\bm{E}_{1}^{\left(0;1,0\right)}$ and $\bm{E}_{2}^{\left(0;1,0\right)}$ cannot be further factorized. That is, they cannot be reduced by product of more than one monomials. In this sense we dub them as being ``unfactorizable'' (or ``prime''). Throughout this paper we shall concentrate on these unfactorizable monomials.  
	\end{enumerate}

Although it is simple for $d=1$, when $d$ becomes large the number of the corresponding monomials (even the unfactorizable ones) becomes huge.
It is helpful to derive a diagrammatic representation of these monomials, which is similar to that employed in the study of tensor networks (see (e.g.,) \cite{Orus:2013kga,Bridgeman:2016dhh} for reviews).
For the covariant derivatives of the scalr field, we denote
\begin{center}
	\includegraphics[scale=0.5]{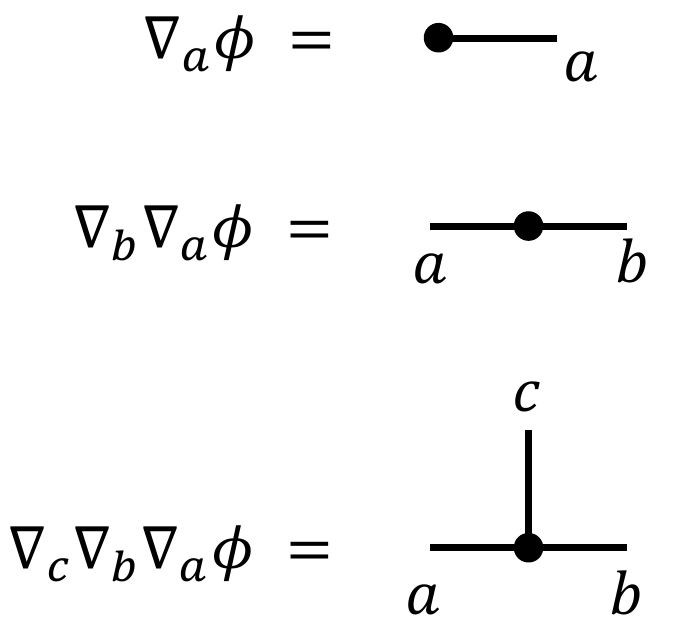}
	\par\end{center}
where a black dot stands for the scalar field $\phi$ and each leg stands for one derivative $\nabla$.
For the Riemann tensor and the Levi-Civita tensor, we denote
\begin{center}
	\includegraphics[scale=0.5]{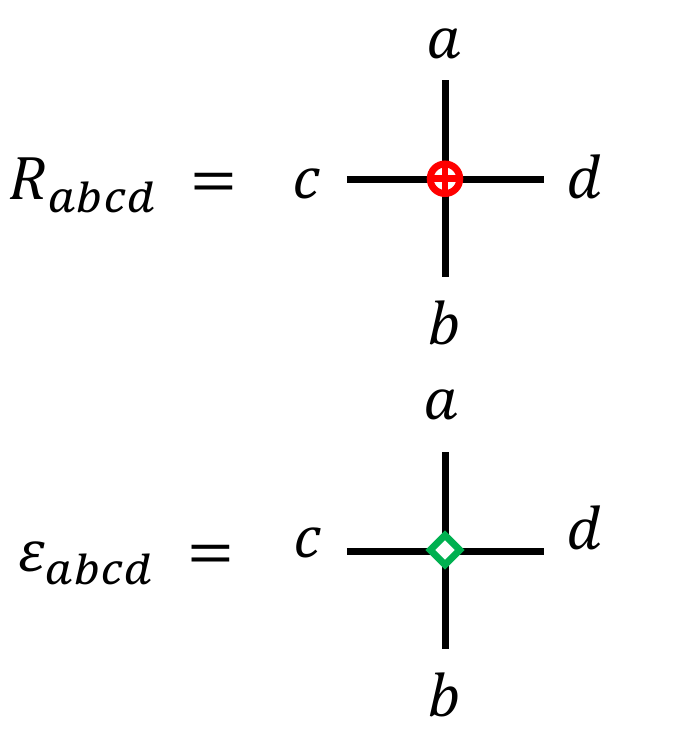}
	\par\end{center}
where each leg stands for one spacetime index. 
In this work, we do not need to consider covariant derivatives of the Riemann tensor, which may be represented by drawing more legs in the above diagram.
Contraction between two indices is thus represented by connecting two legs.

\begin{figure}[H]
	\begin{center}
		\includegraphics[scale=0.5]{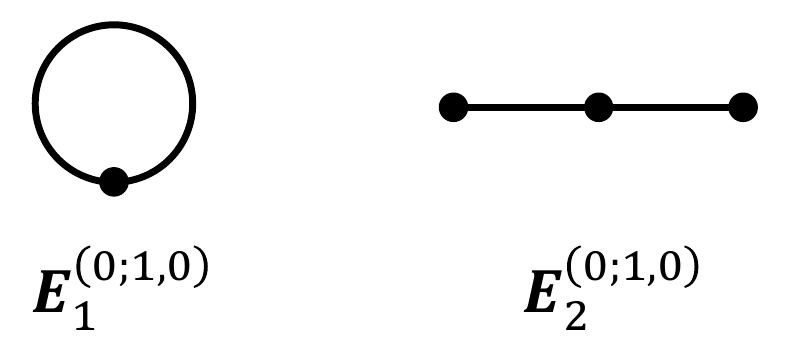}
		\par\end{center}
	\caption{Diagrammatic representation of the monomials with $d=1$.}
	\label{fig:d1}
\end{figure}

According to these simple rules, the diagrammatic representations of the two monomials $\bm{E}_{1}^{\left(0;1,0\right)}$ and $\bm{E}_{1}^{\left(0;1,0\right)}$ are shown in Figure \ref{fig:d1}.
Clearly, we may dub the first and the second diagrams as the ``loop'' and ``tree'' diagrams, respectively.

\section{$d=2$} \label{sec:d2}

In this section, we derive the monomials for $d=2$.
We first derive the unfactorizable monomials for each category according to Table \ref{tab:clsST}.

\subsection{Monomials}

\subsubsection{$(0;2,0)$}

For $(0;2,0)$, there are 2 unfactorizable monomials, which we denote to be
	\begin{eqnarray}
	\bm{E}_{1}^{\left(0;2,0\right)} & \equiv & \frac{1}{\sigma^{2}}\nabla_{a}\nabla_{b}\phi\nabla^{a}\nabla^{b}\phi,\\
	\bm{E}_{2}^{\left(0;2,0\right)} & \equiv & \frac{1}{\sigma^{4}}\nabla^{a}\phi\nabla^{b}\phi\nabla_{c}\nabla_{a}\phi\nabla^{c}\nabla_{b}\phi.
	\end{eqnarray}
In fact, for all the categories $(0;n,0)$ with $n=1,2,3,4$, there are only 2 terms that cannot be factorized.
On the other hand, the products of two monomials of $d=1$ also yield  monomials of $d=2$. 
In our notation, this can be explained as the decomposition 
	\begin{equation}
		\left(0;2,0\right)=\left(0;1,0\right)+\left(0;1,0\right),
	\end{equation} 
which implies that the products of two monomials $\bm{E}^{(0;1,0)}_{m}\bm{E}^{(0;1,0)}_{n}$ are also of the category $(0;2,0)$.
As a result, since there are 2 unfactorizable monomials for $(0;1,0)$, there are  $\frac{2(2+1)}{2}=3$ monomials that are factorizable, i.e., can be expressed in terms of products of multiple unfactorizable monomials. In the case of $(0;2,0)$, these 3 factorizable monomials are
	\begin{equation}
	\left(\bm{E}_{1}^{\left(0;1,0\right)}\right)^{2}, \qquad \bm{E}_{1}^{\left(0;1,0\right)}\bm{E}_{2}^{\left(0;1,0\right)},\qquad \left(\bm{E}_{2}^{\left(0;1,0\right)}\right)^{2}, \label{re020}
	\end{equation}
where $\bm{E}_{1}^{\left(0;1,0\right)}$ and $\bm{E}_{2}^{\left(0;1,0\right)}$ are defined in (\ref{E010_1}) and (\ref{E010_2}), respectively.
For the sake of briefness, here and throughout this paper, we do not write the explicit expressions for these factorizable monomials, which are not important for our purpose.

In total, there are 5 monomials with $(0;2,0)$, which are consistent with the result in \cite{BenAchour:2016fzp} (see eq. (2.7)).

\subsubsection{$(0;0,1)$}

There are 3 terms that are not factorizable, which we denote to be
	\begin{eqnarray}
	\bm{E}_{1}^{\left(0;0,1\right)} & \equiv & \frac{1}{\sigma^{2}}\nabla^{a}\phi\nabla_{a}\square\phi,\\
	\bm{E}_{2}^{\left(0;0,1\right)} & \equiv & \frac{1}{\sigma^{2}}\nabla^{a}\phi\square\nabla_{a}\phi,\\
	\bm{E}_{3}^{\left(0;0,1\right)} & \equiv & \frac{1}{\sigma^{4}}\nabla^{a}\phi\nabla^{b}\phi\nabla^{c}\phi\nabla_{a}\nabla_{b}\nabla_{c}\phi.
	\end{eqnarray}
At the order $d=2$, i.e., if the above monomials enter the Lagrangian linearly with coefficients being functions of $X$, they can be reduced by integrations by parts, as has been shown in Table \ref{tab:clsST}. Here we derive their expressions, which will be used to construct factorizable monomials in $d=3$ and $d=4$.

\subsubsection{$(1;0,0)$}

This category involves the Riemann curvature tensor, which implies the direct coupling between the curvature and derivatives of the scalar field. 
There are 2 unfactorizable monomials, which can be denoted by
	\begin{eqnarray}
	\bm{E}_{1}^{\left(1;0,0\right)} & \equiv & R,\\
	\bm{E}_{2}^{\left(1;0,0\right)} & \equiv & \frac{1}{\sigma^{2}}R_{ab}\nabla^{a}\phi\nabla^{b}\phi.
	\end{eqnarray}

For later convenience, note using the fact that $\left[\square,\nabla_{a}\right]\phi\equiv R_{ab}\nabla^{b}\phi$, we have
	\begin{equation}
	\bm{E}_{2}^{\left(0;0,1\right)}\equiv\bm{E}_{1}^{\left(0;0,1\right)}+\bm{E}_{2}^{\left(1;0,0\right)}.
	\end{equation}
We emphasize that this is an equality, instead of integration by part.
The above fact indicates explicitly that direct coupling between the curvature and derivatives of the scalar field can be viewed as derivatives of the scalar field higher than the second order.

\subsection{Complete basis}

We are now in the position to derive a set of monomials such that any polynomial with $d=2$ can be expressed as a linear combination of the monomials in this set.
For this reason, we refer to this set of monomials as the ``complete basis''.
Since categories belong to the reducible case in Table \ref{tab:clsST} can be suppressed from the beginning, we consider the complete basis for the irreducible case only. 
We emphasize that the ``completeness'' is in the sense of linear combination, by taking into account (anti)symmetries including the Bianchi identities of the Riemann tensor.
At the level of Lagrangian, there might be further reduction after performing the integrations by parts.

Since all the 3 monomials with $(0;0,1)$ are reducible, the complete basis for $d=2$ consists of 4 unfactorizable monomials
	\begin{equation}
	\bm{E}_{1}^{\left(0;2,0\right)},\qquad\bm{E}_{2}^{\left(0;2,0\right)},\qquad\bm{E}_{1}^{\left(1;0,0\right)},\qquad\bm{E}_{2}^{\left(1;0,0\right)}, \label{cb_d2}
	\end{equation}
together with 3 factorizable monomials in (\ref{re020}).
As a quick application, the Horndeski Lagrangian $\mathcal{L}_{4}^{\mathrm{H}}$ in (\ref{LH4}) now can be written briefly as
	\begin{equation}
		\mathcal{L}_{4}^{\text{H}}=X^{2}\left(\bm{E}_{1}^{\left(1;0,0\right)}+4\left(\bm{E}_{1}^{\left(0;1,0\right)}\right)^{2}-4\bm{E}_{1}^{\left(0;2,0\right)}\right).
	\end{equation}

\begin{figure}[H]
		\begin{center}
			\includegraphics[scale=0.5]{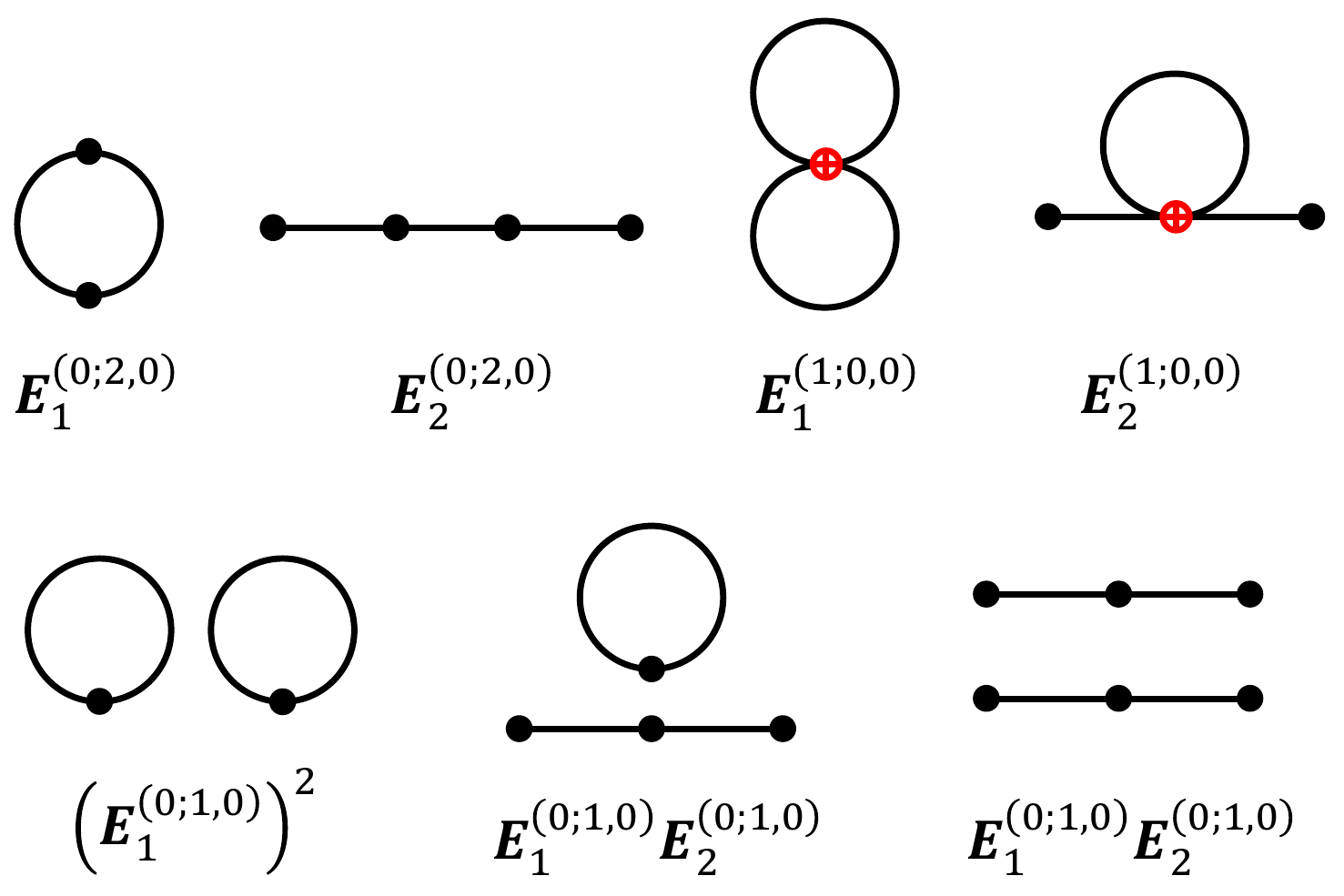}
			\par\end{center}
		\caption{Diagrammatic representation of the basis of monomials for $d=2$.}
		\label{fig:d2}
	\end{figure}

The diagrammatic representation of these 7 monomials are shown in Figure \ref{fig:d2}.
The upper 4 diagrams are the unfactorizable monomials in (\ref{cb_d2}), and the lower 3 diagrams are factorized monomials in (\ref{re020}). Clearly, unfactorizable monomials correspond to ``connect'' diagrams, while factorizable monomials correspond to ``disconnect'' diagrams.

Note all the monomials discussed in the above are parity preserving. It is not possible to construct parity violating monomials for $d=2$.

\section{$d=3$} \label{sec:d3}

In this section, we investigate the monomials for $d=3$.
Again, we first derive all the monomials according to the order in Table \ref{tab:clsST}, then discuss the possible linear dependence among different monomials in order to derive the complete basis of monomials.

\subsection{Monomials}

\subsubsection{$(0;3,0)$}

There are 2 unfactorizable monomials, which we denote to be
	\begin{eqnarray}
	\bm{E}_{1}^{\left(0;3,0\right)} & \equiv & \frac{1}{\sigma^{3}}\nabla_{a}\nabla^{b}\phi\nabla_{b}\nabla^{c}\phi\nabla_{c}\nabla^{a}\phi,\\
	\bm{E}_{2}^{\left(0;3,0\right)} & \equiv & \frac{1}{\sigma^{5}}\nabla^{a}\phi\nabla^{b}\phi\nabla_{a}\nabla_{c}\phi\nabla^{c}\nabla_{d}\phi\nabla^{d}\nabla_{b}\phi.
	\end{eqnarray}
Since
	\begin{eqnarray}
	\left(0;3,0\right) & = & \left(0;1,0\right)+\left(0;1,0\right)+\left(0;1,0\right)\\
	& = & \left(0;1,0\right)+\left(0;2,0\right),
	\end{eqnarray}
there must be another $4+2\times 2=8$ monomials that are can be factorized, which are 
	\begin{eqnarray}
	&  & \left(\bm{E}_{1}^{\left(0;1,0\right)}\right)^{3},\qquad\left(\bm{E}_{1}^{\left(0;1,0\right)}\right)^{2}\bm{E}_{2}^{\left(0;1,0\right)},\qquad\bm{E}_{1}^{\left(0;1,0\right)}\left(\bm{E}_{2}^{\left(0;1,0\right)}\right)^{2},\qquad\left(\bm{E}_{2}^{\left(0;1,0\right)}\right)^{3},\nonumber \\
	&  & \bm{E}_{1}^{\left(0;1,0\right)}\bm{E}_{1}^{\left(0;2,0\right)},\qquad\bm{E}_{1}^{\left(0;1,0\right)}\bm{E}_{2}^{\left(0;2,0\right)},\qquad\bm{E}_{2}^{\left(0;1,0\right)}\bm{E}_{1}^{\left(0;2,0\right)},\qquad\bm{E}_{2}^{\left(0;1,0\right)}\bm{E}_{2}^{\left(0;2,0\right)}. \label{re030}
	\end{eqnarray}
In total, these 10 terms are consistent with the result in \cite{BenAchour:2016fzp} (see eq. (2.8)).

\subsubsection{$(0;1,1)$}

There are 5 monomials that are not factorizable
	\begin{eqnarray}
	\bm{E}_{1}^{\left(0;1,1\right)} & \equiv & \frac{1}{\sigma^{3}}\nabla^{a}\phi\nabla_{a}\nabla^{b}\phi\nabla_{b}\square\phi,\\
	\bm{E}_{2}^{\left(0;1,1\right)} & \equiv & \frac{1}{\sigma^{3}}\nabla^{a}\phi\nabla_{a}\nabla^{b}\phi\square\nabla_{b}\phi,\\
	\bm{E}_{3}^{\left(0;1,1\right)} & \equiv & \frac{1}{\sigma^{3}}\nabla^{a}\phi\nabla^{b}\nabla^{c}\phi\nabla_{a}\nabla_{b}\nabla_{c}\phi,\\
	\bm{E}_{4}^{\left(0;1,1\right)} & \equiv & \frac{1}{\sigma^{3}}\nabla^{a}\phi\nabla^{b}\nabla^{c}\phi\nabla_{b}\nabla_{c}\nabla_{a}\phi,\\
	\bm{E}_{5}^{\left(0;1,1\right)} & \equiv & \frac{1}{\sigma^{5}}\nabla^{a}\phi\nabla^{b}\phi\nabla^{c}\phi\nabla_{a}\nabla^{d}\phi\nabla_{d}\nabla_{b}\nabla_{c}\phi,
	\end{eqnarray}
Since
	\begin{equation}
	\left(0;1,1\right)=\left(0;1,0\right)+\left(0;0,1\right),
	\end{equation}
there are also another $2\times 3=6$ monomials that can be factored.

There is also a single parity-violating term
	\begin{equation}
	\bm{F}_{1}^{\left(0;1,1\right)} \equiv \frac{1}{\sigma^{3}}\varepsilon_{abcd}\nabla^{a}\phi\nabla^{b}\nabla^{f}\phi\nabla^{c}\nabla^{d}\nabla_{f}\phi,
	\end{equation}
which cannot be factorized. 
Clearly, $\bm{F}_{1}^{\left(0;1,1\right)}$ can be recast in terms of curvature tensor due to the antisymmetry of the Levi-Civita tensor and the commutator of two covariant derivatives, and thus is not an independent term. Here we show its expression for notational completeness.

\subsubsection{$(1;1,0)$}

There are 3 unfactorizable monomials
	\begin{eqnarray}
	\bm{E}_{1}^{\left(1;1,0\right)} & \equiv & \frac{1}{\sigma}R_{ab}\nabla^{a}\nabla^{b}\phi,\\
	\bm{E}_{2}^{\left(1;1,0\right)} & \equiv & \frac{1}{\sigma^{3}}R_{abcd}\nabla^{a}\phi\nabla^{c}\phi\nabla^{b}\nabla^{d}\phi,\\
	\bm{E}_{3}^{\left(1;1,0\right)} & \equiv & \frac{1}{\sigma^{3}}R_{ab}\nabla^{a}\phi\nabla^{c}\phi\nabla^{b}\nabla_{c}\phi,
	\end{eqnarray}
Since
	\begin{equation}
	\left(1;1,0\right)=\left(1;0,0\right)+\left(0;1,0\right),
	\end{equation}
there are another $2\times 2=4$ factorizable monomials.

In the case of parity violation, there is a single term
	\begin{equation}
	\bm{F}_{1}^{\left(1;1,0\right)}\equiv\frac{1}{\sigma^{3}}\varepsilon_{abcd}R_{ef}^{\phantom{ef}cd}\nabla^{a}\phi\nabla^{e}\phi\nabla^{b}\nabla^{f}\phi,
	\end{equation}
which has also been considered in \cite{Crisostomi:2017ugk} (see eq.(3.12)).

\subsection{Complete basis}

\begin{figure}[H]
		\begin{center}
			\includegraphics[scale=0.5]{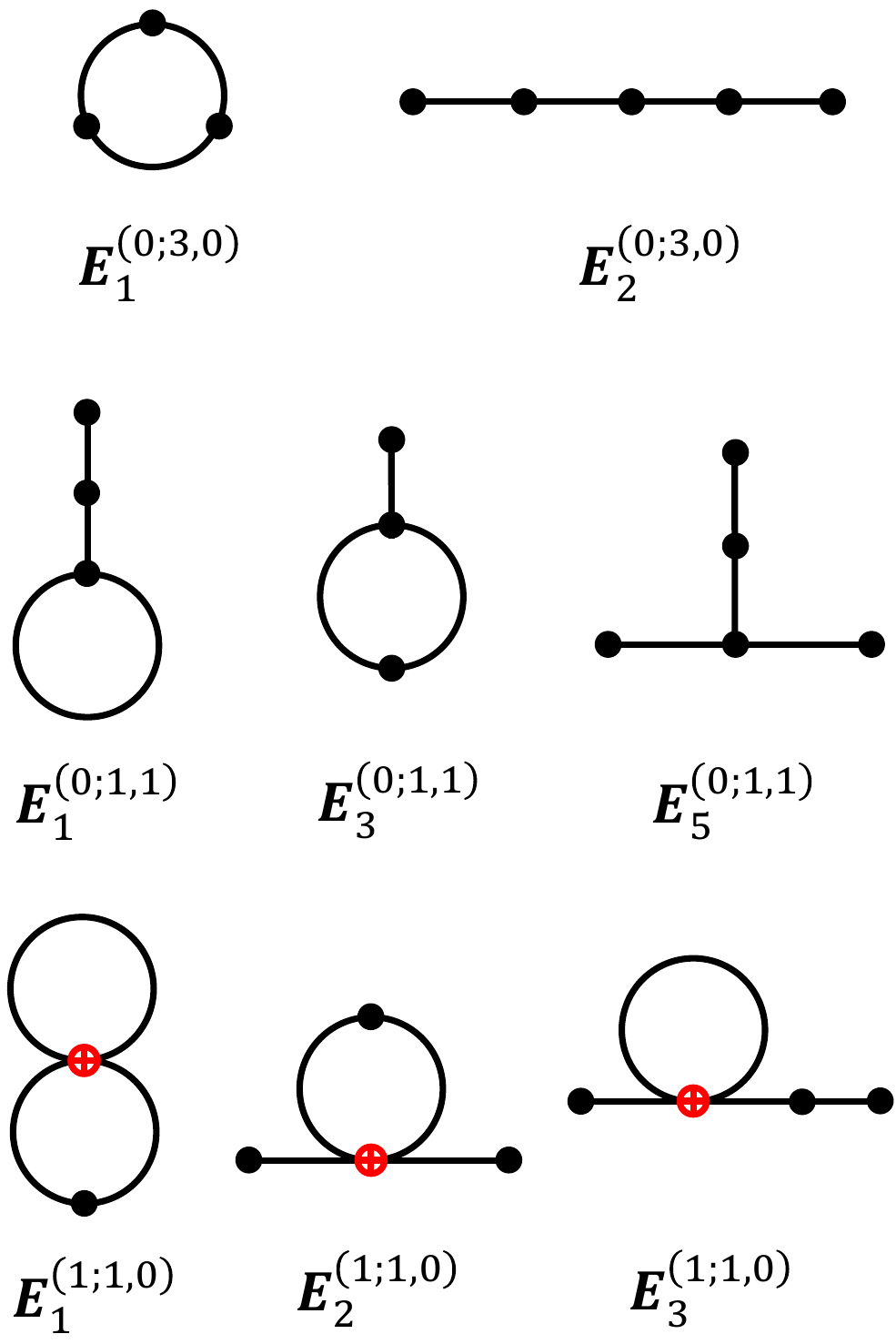}
			\par\end{center}
		\caption{Diagrammatic representation of the 8 parity preserving unfactorizable monomials for $d=3$.}
		\label{fig:d3}
	\end{figure}

By using the antisymmetry of the Levi-Civita tensor and the fact that Riemann tensor is the commutator of two covariant derivatives, we get the following linear dependence among different monomials:
	\begin{eqnarray}
	\bm{E}_{2}^{\left(0;1,1\right)} & \equiv & \bm{E}_{1}^{\left(0;1,1\right)}+\bm{E}_{3}^{(1;1,0)},\\
	\bm{E}_{4}^{\left(0;1,1\right)} & \equiv & \bm{E}_{3}^{\left(0;1,1\right)}+\bm{E}_{2}^{(1;1,0)}.
	\end{eqnarray}
As a result, the complete basis for $d=3$ consists of 8 irreducible monomials
	\begin{eqnarray}
	&  & \bm{E}_{1}^{\left(0;3,0\right)},\qquad\bm{E}_{2}^{\left(0;3,0\right)},\nonumber \\
	&  & \bm{E}_{1}^{\left(0;1,1\right)},\qquad\bm{E}_{3}^{\left(0;1,1\right)},\qquad\bm{E}_{5}^{\left(0;1,1\right)},\nonumber \\
	&  & \bm{E}_{1}^{\left(1;1,0\right)},\qquad\bm{E}_{2}^{\left(1;1,0\right)},\qquad\bm{E}_{3}^{\left(1;1,0\right)}, \label{cb_d3}
	\end{eqnarray}
together with 8+6+4=18 factorizable monomials.
The diagrammatic representation of the 8 unfactorizable monomials are shown in Figure \ref{fig:d3}.

\begin{figure}[H]
		\begin{center}
			\includegraphics[scale=0.5]{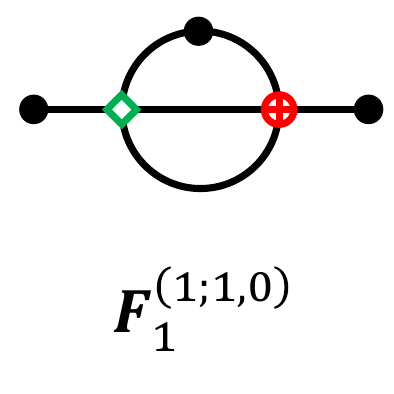}
			\par\end{center}
		\caption{Diagrammatic representation of the single parity violating monomial for $d=3$.}
		\label{fig:d3p}
	\end{figure}

In the case of parity violation, we have
	\begin{equation}
	\bm{F}_{1}^{\left(0;1,1\right)}\equiv-\frac{1}{2}\bm{F}_{1}^{(1;1,0)}, \label{F011_F110}
	\end{equation}
thus $\bm{F}_{1}^{(1;1,0)}$ is the single independent parity-violating term for $d=3$.
The diagram for $\bm{F}_{1}^{(1;1,0)}$ is shown in Figure \ref{fig:d3p}. 

\section{$d=4$} \label{sec:d4}

In this section we investigate the monomials for $d=4$, which are the most involved ones in this work.

\subsection{Monomials}

\subsubsection{$(0;4,0)$}

There are 2 unfactorizable monomials, which are
	\begin{eqnarray}
	\bm{E}_{1}^{\left(0;4,0\right)} & \equiv & \frac{1}{\sigma^{4}}\nabla^{a}\nabla^{b}\phi\nabla^{c}\nabla_{b}\phi\nabla_{c}\nabla_{d}\phi\nabla^{d}\nabla_{a}\phi,\\
	\bm{E}_{2}^{\left(0;4,0\right)} & \equiv & \frac{1}{\sigma^{6}}\nabla^{a}\phi\nabla^{b}\phi\nabla_{a}\nabla_{c}\phi\nabla^{c}\nabla_{d}\phi\nabla^{d}\nabla^{e}\phi\nabla_{e}\nabla_{b}\phi.
	\end{eqnarray}
Since
	\begin{eqnarray}
	\left(0;4,0\right) & = & \left(0;1,0\right)+\left(0;1,0\right)+\left(0;1,0\right)+\left(0;1,0\right)\\
	& = & \left(0;1,0\right)+\left(0;1,0\right)+\left(0;2,0\right)\\
	& = & \left(0;2,0\right)+\left(0;2,0\right)\\
	& = & \left(0;1,0\right)+\left(0;3,0\right),
	\end{eqnarray}
there are $5+3 \times 2+3+2 \times 2=18$ monomials that can be factorized, which are 
	\begin{eqnarray}
	&  & \left(\bm{E}_{1}^{(0;1,0)}\right)^{4},\qquad\left(\bm{E}_{1}^{(0;1,0)}\right)^{3}\bm{E}_{2}^{(0;1,0)},\qquad\left(\bm{E}_{1}^{(0;1,0)}\right)^{2}\left(\bm{E}_{2}^{(0;1,0)}\right)^{2},\qquad\bm{E}_{1}^{(0;1,0)}\left(\bm{E}_{2}^{(0;1,0)}\right)^{3},\nonumber \\
	&  & \left(\bm{E}_{2}^{(0;1,0)}\right)^{4},\qquad\left(\bm{E}_{1}^{(0;1,0)}\right)^{2}\bm{E}_{1}^{(0;2,0)},\qquad\bm{E}_{1}^{(0;1,0)}\bm{E}_{2}^{(0;1,0)}\bm{E}_{1}^{(0;2,0)},\qquad\left(\bm{E}_{2}^{(0;1,0)}\right)^{2}\bm{E}_{1}^{(0;2,0)},\nonumber \\
	&  & \left(\bm{E}_{1}^{(0;1,0)}\right)^{2}\bm{E}_{2}^{(0;2,0)},\qquad\bm{E}_{1}^{(0;1,0)}\bm{E}_{2}^{(0;1,0)}\bm{E}_{2}^{(0;2,0)},\qquad\left(\bm{E}_{2}^{(0;1,0)}\right)^{2}\bm{E}_{2}^{(0;2,0)},\qquad\left(\bm{E}_{1}^{(0;2,0)}\right)^{2},\nonumber \\
	&  & \left(\bm{E}_{2}^{(0;2,0)}\right)^{2},\qquad\bm{E}_{1}^{(0;2,0)}\bm{E}_{2}^{(0;2,0)},\qquad\bm{E}_{1}^{(0;1,0)}\bm{E}_{1}^{(0;3,0)},\qquad\bm{E}_{1}^{(0;1,0)}\bm{E}_{2}^{(0;3,0)},\nonumber \\
	&  & \bm{E}_{2}^{(0;1,0)}\bm{E}_{1}^{(0;3,0)},\qquad\bm{E}_{2}^{(0;1,0)}\bm{E}_{2}^{(0;3,0)}.
	\end{eqnarray}
As a result, there are in total 20 monomials for $(0;4,0)$.

There is no parity violating terms built of the second order derivatives only.

\subsubsection{$(0;2,1)$}

There are 9 unfactorizable monomials 
	\begin{eqnarray}
	\bm{E}_{1}^{\left(0;2,1\right)} & \equiv & \frac{1}{\sigma^{4}}\nabla^{a}\phi\nabla_{a}\nabla^{b}\phi\nabla^{c}\nabla_{b}\phi\nabla_{c}\square\phi,\\
	\bm{E}_{2}^{\left(0;2,1\right)} & \equiv & \frac{1}{\sigma^{4}}\nabla^{a}\phi\nabla_{a}\nabla^{b}\phi\nabla^{c}\nabla_{b}\phi\square\nabla_{c}\phi,\\
	\bm{E}_{3}^{\left(0;2,1\right)} & \equiv & \frac{1}{\sigma^{4}}\nabla^{a}\phi\nabla_{a}\nabla^{b}\phi\nabla^{c}\nabla^{d}\phi\nabla_{b}\nabla_{c}\nabla_{d}\phi,\\
	\bm{E}_{4}^{\left(0;2,1\right)} & \equiv & \frac{1}{\sigma^{4}}\nabla^{a}\phi\nabla_{a}\nabla^{b}\phi\nabla^{c}\nabla^{d}\phi\nabla_{c}\nabla_{d}\nabla_{b}\phi,\\
	\bm{E}_{5}^{\left(0;2,1\right)} & \equiv & \frac{1}{\sigma^{4}}\nabla^{a}\phi\nabla^{b}\nabla^{c}\phi\nabla^{d}\nabla_{c}\phi\nabla_{a}\nabla_{b}\nabla_{d}\phi,\\
	\bm{E}_{6}^{\left(0;2,1\right)} & \equiv & \frac{1}{\sigma^{4}}\nabla^{a}\phi\nabla^{b}\nabla^{c}\phi\nabla^{d}\nabla_{c}\phi\nabla_{b}\nabla_{d}\nabla_{a}\phi,\\
	\bm{E}_{7}^{\left(0;2,1\right)} & \equiv & \frac{1}{\sigma^{6}}\nabla^{a}\phi\nabla^{b}\phi\nabla^{c}\phi\nabla^{d}\nabla_{a}\phi\nabla^{e}\nabla_{d}\phi\nabla_{e}\nabla_{b}\nabla_{c}\phi,\\
	\bm{E}_{8}^{\left(0;2,1\right)} & \equiv & \frac{1}{\sigma^{6}}\nabla^{a}\phi\nabla^{b}\phi\nabla^{c}\phi\nabla^{d}\nabla_{a}\phi\nabla^{e}\nabla_{b}\phi\nabla_{c}\nabla_{d}\nabla_{e}\phi,\\
	\bm{E}_{9}^{\left(0;2,1\right)} & \equiv & \frac{1}{\sigma^{6}}\nabla^{a}\phi\nabla^{b}\phi\nabla^{c}\phi\nabla^{d}\nabla_{a}\phi\nabla^{e}\nabla_{b}\phi\nabla_{d}\nabla_{e}\nabla_{c}\phi,
	\end{eqnarray}
Since
	\begin{eqnarray}
	\left(0;2,1\right) & = & \left(0;1,0\right)+\left(0;1,0\right)+\left(0;0,1\right)\\
	& = & \left(0;2,0\right)+\left(0;0,1\right)\\
	& = & \left(0;1,0\right)+\left(0;1,1\right),
	\end{eqnarray}
there are another $3\times 3+2\times 3+2\times 5=25$ monomials that are factorizable. Here and in the rest part of this section, we do not show the explicit expressions of these factorizable monomials due to their length, which can be read straightforwardly.

There are also parity violating contractions.
There are 6 unfactorizable monomials
	\begin{eqnarray}
	\bm{F}_{1}^{\left(0;2,1\right)} & \equiv & \frac{1}{\sigma^{4}}\varepsilon_{abcd}\nabla^{a}\phi\nabla^{b}\nabla^{e}\phi\nabla^{c}\nabla^{f}\phi\nabla_{e}\nabla_{f}\nabla^{d}\phi,\\
	\bm{F}_{2}^{\left(0;2,1\right)} & \equiv & \frac{1}{\sigma^{4}}\varepsilon_{abcd}\nabla^{a}\phi\nabla^{e}\nabla^{f}\phi\nabla^{b}\nabla_{e}\phi\nabla^{c}\nabla^{d}\nabla_{f}\phi,\\
	\bm{F}_{3}^{\left(0;2,1\right)} & \equiv & \frac{1}{\sigma^{4}}\varepsilon_{abcd}\nabla^{e}\phi\nabla^{a}\nabla^{f}\phi\nabla^{b}\nabla_{e}\phi\nabla^{c}\nabla^{d}\nabla_{f}\phi,\\
	\bm{F}_{4}^{\left(0;2,1\right)} & \equiv & \frac{1}{\sigma^{6}}\varepsilon_{abcd}\nabla^{m}\phi\nabla^{e}\phi\nabla^{a}\phi\nabla^{f}\nabla_{m}\phi\nabla^{b}\nabla_{e}\phi\nabla^{c}\nabla^{d}\nabla_{f}\phi,\\
	\bm{F}_{5}^{\left(0;2,1\right)} & \equiv & \frac{1}{\sigma^{6}}\varepsilon_{abcd}\nabla^{e}\phi\nabla^{f}\phi\nabla^{a}\phi\nabla^{b}\nabla_{e}\phi\nabla^{c}\nabla^{m}\phi\nabla_{f}\nabla_{m}\nabla^{d}\phi,\\
	\bm{F}_{6}^{\left(0;2,1\right)} & \equiv & \frac{1}{\sigma^{6}}\varepsilon_{abcd}\nabla^{e}\phi\nabla^{f}\phi\nabla^{a}\phi\nabla^{b}\nabla_{e}\phi\nabla^{c}\nabla^{m}\phi\nabla_{m}\nabla_{f}\nabla^{d}\phi.
	\end{eqnarray}
In the case of parity violation, since $\left(0;2,1\right)=\left(0;1,0\right)+\left(0;1,1\right)$, there are another 2 terms that are factorizable, i.e.,
	\begin{equation}
		\bm{E}_{1}^{(0;1,0)} \bm{F}_{1}^{(0;1,1)}, \qquad \bm{E}_{2}^{(0;1,0)} \bm{F}_{1}^{(0;1,1)},
	\end{equation}
or equivalently (recall eq. (\ref{F011_F110}))
	\begin{equation}
	\bm{E}_{1}^{(0;1,0)} \bm{F}_{1}^{(1;1,0)}, \qquad \bm{E}_{2}^{(0;1,0)} \bm{F}_{1}^{(1;1,0)}.
	\end{equation}

\subsubsection{$(0;0,2)$}

There are 11 unfactorizable monomials
	\begin{eqnarray}
	\bm{E}_{1}^{\left(0;0,2\right)} & \equiv & \frac{1}{\sigma^{2}}\nabla_{a}\square\phi\nabla^{a}\square\phi,\\
	\bm{E}_{2}^{\left(0;0,2\right)} & \equiv & \frac{1}{\sigma^{2}}\square\nabla^{a}\phi\square\nabla_{a}\phi,\\
	\bm{E}_{3}^{\left(0;0,2\right)} & \equiv & \frac{1}{\sigma^{2}}\nabla^{a}\square\phi\square\nabla_{a}\phi,\\
	\bm{E}_{4}^{\left(0;0,2\right)} & \equiv & \frac{1}{\sigma^{2}}\nabla_{a}\nabla_{b}\nabla_{c}\phi\nabla^{a}\nabla^{b}\nabla^{c}\phi,\\
	\bm{E}_{5}^{\left(0;0,2\right)} & \equiv & \frac{1}{\sigma^{2}}\nabla_{a}\nabla_{b}\nabla_{c}\phi\nabla^{b}\nabla^{c}\nabla^{a}\phi,\\
	\bm{E}_{6}^{\left(0;0,2\right)} & \equiv & \frac{1}{\sigma^{4}}\nabla^{a}\phi\nabla^{b}\phi\nabla^{c}\nabla_{a}\nabla_{b}\phi\nabla_{c}\square\phi,\\
	\bm{E}_{7}^{\left(0;0,2\right)} & \equiv & \frac{1}{\sigma^{4}}\nabla^{a}\phi\nabla^{b}\phi\nabla^{c}\nabla_{a}\nabla_{b}\phi\square\nabla_{c}\phi,\\
	\bm{E}_{8}^{\left(0;0,2\right)} & \equiv & \frac{1}{\sigma^{4}}\nabla^{a}\phi\nabla^{b}\phi\nabla_{c}\nabla_{d}\nabla_{a}\phi\nabla^{c}\nabla^{d}\nabla_{b}\phi,\\
	\bm{E}_{9}^{\left(0;0,2\right)} & \equiv & \frac{1}{\sigma^{4}}\nabla^{a}\phi\nabla^{b}\phi\nabla_{a}\nabla_{c}\nabla_{d}\phi\nabla_{b}\nabla^{c}\nabla^{d}\phi,\\
	\bm{E}_{10}^{\left(0;0,2\right)} & \equiv & \frac{1}{\sigma^{4}}\nabla^{a}\phi\nabla^{b}\phi\nabla_{a}\nabla_{c}\nabla_{d}\phi\nabla^{c}\nabla^{d}\nabla_{b}\phi,\\
	\bm{E}_{11}^{\left(0;0,2\right)} & \equiv & \frac{1}{\sigma^{6}}\nabla^{a}\phi\nabla^{b}\phi\nabla^{c}\phi\nabla^{d}\phi\nabla^{e}\nabla_{a}\nabla_{b}\phi\nabla_{e}\nabla_{c}\nabla_{d}\phi.
	\end{eqnarray}
Since 
	\begin{equation}
	\left(0;0,2\right)=\left(0;0,1\right)+\left(0;0,1\right),
	\end{equation}
there are another $\frac{3(3+1)}{2}=6$ monomials that are factorizable.

In the case of parity violation, there are also 3 monomials
	\begin{eqnarray}
	\bm{F}_{1}^{\left(0;0,2\right)} & \equiv & \frac{1}{\sigma^{2}}\varepsilon_{abcd}\nabla^{a}\nabla^{b}\nabla^{e}\phi\nabla^{c}\nabla^{d}\nabla_{e}\phi,\\
	\bm{F}_{2}^{\left(0;0,2\right)} & \equiv & \frac{1}{\sigma^{4}}\varepsilon_{abcd}\nabla^{a}\phi\nabla^{e}\phi\nabla^{b}\nabla^{f}\nabla_{e}\phi\nabla^{c}\nabla^{d}\nabla_{f}\phi,\\
	\bm{F}_{3}^{\left(0;0,2\right)} & \equiv & \frac{1}{\sigma^{4}}\varepsilon_{abcd}\nabla^{a}\phi\nabla^{e}\phi\nabla_{e}\nabla^{f}\nabla^{b}\phi\nabla^{c}\nabla^{d}\nabla_{f}\phi,
	\end{eqnarray}
which are not factorizable.
There is no factorizable monomials with parity violating, since although $(0;0,2) \sim (0;0,1)+(0;0,1)$, there is no parity violating monomials of $(0;0,1)$.

\subsubsection{$(1;2,0)$}

There are 7 unfactorizable monomials
	\begin{eqnarray}
	\bm{E}_{1}^{\left(1;2,0\right)} & \equiv & \frac{1}{\sigma^{2}}R_{abcd}\,\nabla^{a}\nabla^{c}\phi\nabla^{b}\nabla^{d}\phi,\\
	\bm{E}_{2}^{\left(1;2,0\right)} & \equiv & \frac{1}{\sigma^{2}}R^{ab}\,\nabla_{a}\nabla^{c}\phi\nabla_{b}\nabla_{c}\phi,\\
	\bm{E}_{3}^{\left(1;2,0\right)} & \equiv & \frac{1}{\sigma^{4}}R_{abcd}\nabla^{a}\phi\nabla^{c}\phi\nabla^{b}\nabla^{e}\phi\nabla^{d}\nabla_{e}\phi,\\
	\bm{E}_{4}^{\left(1;2,0\right)} & \equiv & \frac{1}{\sigma^{4}}R_{abcd}\nabla^{a}\phi\nabla^{e}\phi\nabla^{c}\nabla_{e}\phi\nabla^{b}\nabla^{d}\phi,\\
	\bm{E}_{5}^{\left(1;2,0\right)} & \equiv & \frac{1}{\sigma^{4}}R^{ab}\nabla^{c}\phi\nabla^{d}\phi\nabla_{a}\nabla_{c}\phi\nabla_{b}\nabla_{d}\phi,\\
	\bm{E}_{6}^{\left(1;2,0\right)} & \equiv & \frac{1}{\sigma^{4}}R^{ab}\nabla_{a}\phi\nabla^{c}\phi\nabla_{b}\nabla_{d}\phi\nabla_{c}\nabla^{d}\phi,\\
	\bm{E}_{7}^{\left(1;2,0\right)} & \equiv & \frac{1}{\sigma^{6}}R_{abcd}\nabla^{a}\phi\nabla^{c}\phi\nabla^{f}\phi\nabla^{e}\phi\nabla^{b}\nabla_{f}\phi\nabla^{d}\nabla_{e}\phi.
	\end{eqnarray}
Since 
	\begin{eqnarray}
	\left(1;2,0\right) & = & \left(1;0,0\right)+\left(0;2,0\right)\\
	& = & \left(1;0,0\right)+\left(0;1,0\right)+\left(0;1,0\right)\\
	& = & \left(1;1,0\right)+\left(0;1,0\right)
	\end{eqnarray}
there are another $2\times 2+2\times 3+3\times 2=16$ factorizable monomials, of which the expressions can be read straightforwardly.

In the case of parity violation, there are 8 unfactorizable monomials
	\begin{eqnarray}
	\bm{F}_{1}^{\left(1;2,0\right)} & \equiv & \frac{1}{\sigma^{2}}\varepsilon_{abcd} R_{ef}^{\phantom{ef}cd}\nabla^{a}\nabla^{e}\phi\nabla^{b}\nabla^{f}\phi,\\
	\bm{F}_{2}^{\left(1;2,0\right)} & \equiv & \frac{1}{\sigma^{4}}\varepsilon_{abcd} R_{ef}^{\phantom{ef}cd}\nabla^{a}\phi\nabla^{e}\phi\nabla^{b}\nabla_{m}\phi\nabla^{f}\nabla^{m}\phi,\\
	\bm{F}_{3}^{\left(1;2,0\right)} & \equiv & \frac{1}{\sigma^{4}}\varepsilon_{abcd} R_{ef}^{\phantom{ef}cd}\nabla^{e}\phi\nabla^{m}\phi\nabla^{a}\nabla_{m}\phi\nabla^{b}\nabla^{f}\phi,\\
	\bm{F}_{4}^{\left(1;2,0\right)} & \equiv & \frac{1}{\sigma^{4}}\varepsilon_{abcd} R_{ef}^{\phantom{ef}cd}\nabla^{a}\phi\nabla^{m}\phi\nabla^{b}\nabla^{e}\phi\nabla^{f}\nabla_{m}\phi,\\
	\bm{F}_{5}^{\left(1;2,0\right)} & \equiv & \frac{1}{\sigma^{4}}\varepsilon_{abcd} R_{ef}^{\phantom{ef}cm}\nabla^{a}\phi\nabla^{e}\phi\nabla^{b}\nabla^{f}\phi\nabla^{d}\nabla_{m}\phi,\\
	\bm{F}_{6}^{\left(1;2,0\right)} & \equiv & \frac{1}{\sigma^{4}}\varepsilon_{abcd} R^{ae}\nabla^{b}\phi\nabla^{f}\phi\nabla^{c}\nabla_{e}\phi\nabla^{d}\nabla_{f}\phi,\\
	\bm{F}_{7}^{\left(1;2,0\right)} & \equiv & \frac{1}{\sigma^{6}}\varepsilon_{abcd} R_{ef}^{\phantom{ef}cd}\nabla^{m}\phi\nabla^{n}\phi\nabla^{e}\phi\nabla^{a}\phi\nabla^{f}\nabla_{m}\phi\nabla^{b}\nabla_{n}\phi,\\
	\bm{F}_{8}^{\left(1;2,0\right)} & \equiv & \frac{1}{\sigma^{6}}\varepsilon_{abcd} R_{ef}^{\phantom{ef}cm}\nabla^{a}\phi\nabla^{e}\phi\nabla_{m}\phi\nabla^{n}\phi\nabla^{b}\nabla_{n}\phi\nabla^{d}\nabla^{f}\phi.
	\end{eqnarray}
Some of these monomials were considered in \cite{Crisostomi:2017ugk} (see eq. (3.13)).
Since $\left(1;2,0\right)=\left(1;1,0\right)+\left(0;1,0\right)$, there are 2 monomials that are factorizable.

\subsubsection{$(2;0,0)$}

There are 6 unfactorizable monomials
	\begin{eqnarray}
	\bm{E}_{1}^{\left(2;0,0\right)} & \equiv & R_{abcd}R^{abcd},\\
	\bm{E}_{2}^{\left(2;0,0\right)} & \equiv & R_{ab}R^{ab},\\
	\bm{E}_{3}^{\left(2;0,0\right)} & \equiv & \frac{1}{\sigma^{2}}R_{a}^{\phantom{a}cde}R_{bcde}\,\nabla^{a}\phi\nabla^{b}\phi,\\
	\bm{E}_{4}^{\left(2;0,0\right)} & \equiv & \frac{1}{\sigma^{2}}R_{acbd}R^{ab}\,\nabla^{c}\phi\nabla^{d}\phi,\\
	\bm{E}_{5}^{\left(2;0,0\right)} & \equiv & \frac{1}{\sigma^{2}}R_{ac}R_{\phantom{c}b}^{c}\,\nabla^{a}\phi\nabla^{b}\phi,\\
	\bm{E}_{6}^{\left(2;0,0\right)} & \equiv & \frac{1}{\sigma^{4}}R_{a\phantom{e}b}^{\phantom{a}e\phantom{b}f}R_{cedf}\nabla^{a}\phi\nabla^{b}\phi\nabla^{c}\phi\nabla^{d}\phi.
	\end{eqnarray}
Since
	\begin{equation}
	\left(2;0,0\right)=\left(1;0,0\right)+\left(1;0,0\right),
	\end{equation}
there are another $\frac{2(2+1)}{2}=3$ factorizable monomials.

In the case of parity violation, there are 5 unfactorizable monomials 
	\begin{eqnarray}
	\bm{F}_{1}^{\left(2;0,0\right)} & \equiv & \varepsilon_{abcd}R_{ef}^{\phantom{ef}cd}R^{abef},\\
	\bm{F}_{2}^{\left(2;0,0\right)} & \equiv & \frac{1}{\sigma^{2}}\,\varepsilon_{abcd}R_{ef}^{\phantom{ef}cd}R_{\phantom{abf}m}^{abf}\nabla^{e}\phi\nabla^{m}\phi,\\
	\bm{F}_{3}^{\left(2;0,0\right)} & \equiv & \frac{1}{\sigma^{2}}\,\varepsilon_{abcd}R_{ef}^{\phantom{ef}cd}R_{\phantom{efb}m}^{efa}\nabla^{b}\phi\nabla^{m}\phi,\\
	\bm{F}_{4}^{\left(2;0,0\right)} & \equiv & \frac{1}{\sigma^{2}}\,\varepsilon_{abcd}R_{ef}^{\phantom{ef}cd}R^{ae}\nabla^{b}\phi\nabla^{f}\phi,\\
	\bm{F}_{5}^{\left(2;0,0\right)} & \equiv & \frac{1}{\sigma^{4}}\,\varepsilon_{abcd}R_{ef}^{\phantom{ef}cd}R^{amen}\nabla^{b}\phi\nabla^{f}\phi\nabla_{m}\phi\nabla_{n}\phi.
	\end{eqnarray}
Some of these terms were considered in  \cite{Crisostomi:2017ugk} (see eq. (3.1)). There is no factorizable monomial since although $\left(2;0,0\right)=\left(1;0,0\right)+\left(1;0,0\right)$,
there is no parity violating term of $\left(1;0,0\right)$.

\subsubsection{$(1;0,1)$}

There are 8 unfactorizable monomials
	\begin{eqnarray}
	\bm{E}_{1}^{\left(1;0,1\right)} & \equiv & \frac{1}{\sigma^{2}}R_{abcd}\nabla^{a}\phi\nabla^{c}\nabla^{d}\nabla^{b}\phi,\\
	\bm{E}_{2}^{\left(1;0,1\right)} & \equiv & \frac{1}{\sigma^{2}}R^{ab}\nabla_{a}\phi\nabla_{b}\square\phi,\\
	\bm{E}_{3}^{\left(1;0,1\right)} & \equiv & \frac{1}{\sigma^{2}}R^{ab}\nabla_{a}\phi\square\nabla_{b}\phi,\\
	\bm{E}_{4}^{\left(1;0,1\right)} & \equiv & \frac{1}{\sigma^{2}}R^{ab}\nabla^{c}\phi\nabla_{c}\nabla_{a}\nabla_{b}\phi,\\
	\bm{E}_{5}^{\left(1;0,1\right)} & \equiv & \frac{1}{\sigma^{2}}R^{ab}\nabla^{c}\phi\nabla_{a}\nabla_{b}\nabla_{c}\phi,\\
	\bm{E}_{6}^{\left(1;0,1\right)} & \equiv & \frac{1}{\sigma^{4}}R_{abcd}\nabla^{a}\phi\nabla^{c}\phi\nabla^{e}\phi\nabla_{e}\nabla^{b}\nabla^{d}\phi,\\
	\bm{E}_{7}^{\left(1;0,1\right)} & \equiv & \frac{1}{\sigma^{4}}R_{abcd}\nabla^{a}\phi\nabla^{c}\phi\nabla^{e}\phi\nabla^{b}\nabla^{d}\nabla_{e}\phi,\\
	\bm{E}_{8}^{\left(1;0,1\right)} & \equiv & \frac{1}{\sigma^{4}}R^{ab}\nabla_{a}\phi\nabla^{c}\phi\nabla^{d}\phi\nabla_{b}\nabla_{c}\nabla_{d}\phi.
	\end{eqnarray}
Since
	\begin{equation}
	\left(1;0,1\right)=\left(1;0,0\right)+\left(0;0,1\right),
	\end{equation}
there are another $2\times 3=6$ factorizable monomials.

In the case of parity violation, there are 6 unfactorizable monomials 
	\begin{eqnarray}
	\bm{F}_{1}^{\left(1;0,1\right)} & \equiv & \frac{1}{\sigma^{2}}\varepsilon_{abcd}R_{ef}^{\phantom{ef}cd}\nabla^{e}\phi\nabla^{a}\nabla^{b}\nabla^{f}\phi,\\
	\bm{F}_{2}^{\left(1;0,1\right)} & \equiv & \frac{1}{\sigma^{2}}\varepsilon_{abcd}R_{ef}^{\phantom{ef}cd}\nabla^{a}\phi\nabla^{e}\nabla^{f}\nabla^{b}\phi,\\
	\bm{F}_{3}^{\left(1;0,1\right)} & \equiv & \frac{1}{\sigma^{2}}\varepsilon_{abcd}R^{ae}\nabla^{b}\phi\nabla^{c}\nabla^{d}\nabla_{e}\phi,\\
	\bm{F}_{4}^{\left(1;0,1\right)} & \equiv & \frac{1}{\sigma^{4}}\varepsilon_{abcd}R_{ef}^{\phantom{ef}cd}\nabla^{a}\phi\nabla^{e}\phi\nabla^{m}\phi\nabla^{b}\nabla^{f}\nabla_{m}\phi,\\
	\bm{F}_{5}^{\left(1;0,1\right)} & \equiv & \frac{1}{\sigma^{4}}\varepsilon_{abcd}R_{ef}^{\phantom{ef}cd}\nabla^{a}\phi\nabla^{e}\phi\nabla^{m}\phi\nabla_{m}\nabla^{b}\nabla^{f}\phi,\\
	\bm{F}_{6}^{\left(1;0,1\right)} & \equiv & \frac{1}{\sigma^{4}}\varepsilon_{abcd}R_{ef}^{\phantom{ef}cm}\nabla^{a}\phi\nabla^{e}\phi\nabla_{m}\phi\nabla^{d}\nabla^{b}\nabla^{f}\phi.
	\end{eqnarray}
There is no factorizable monomial with $(1;0,1)$ since although $\left(1;0,1\right)=\left(1;0,0\right)+\left(0;0,1\right)$, there is no parity violating monomial with neither $(1;0,0)$ nor $(0;0,1)$.

\subsection{Complete basis}

\begin{figure}[H]
	\begin{center}
		\includegraphics[scale=0.5]{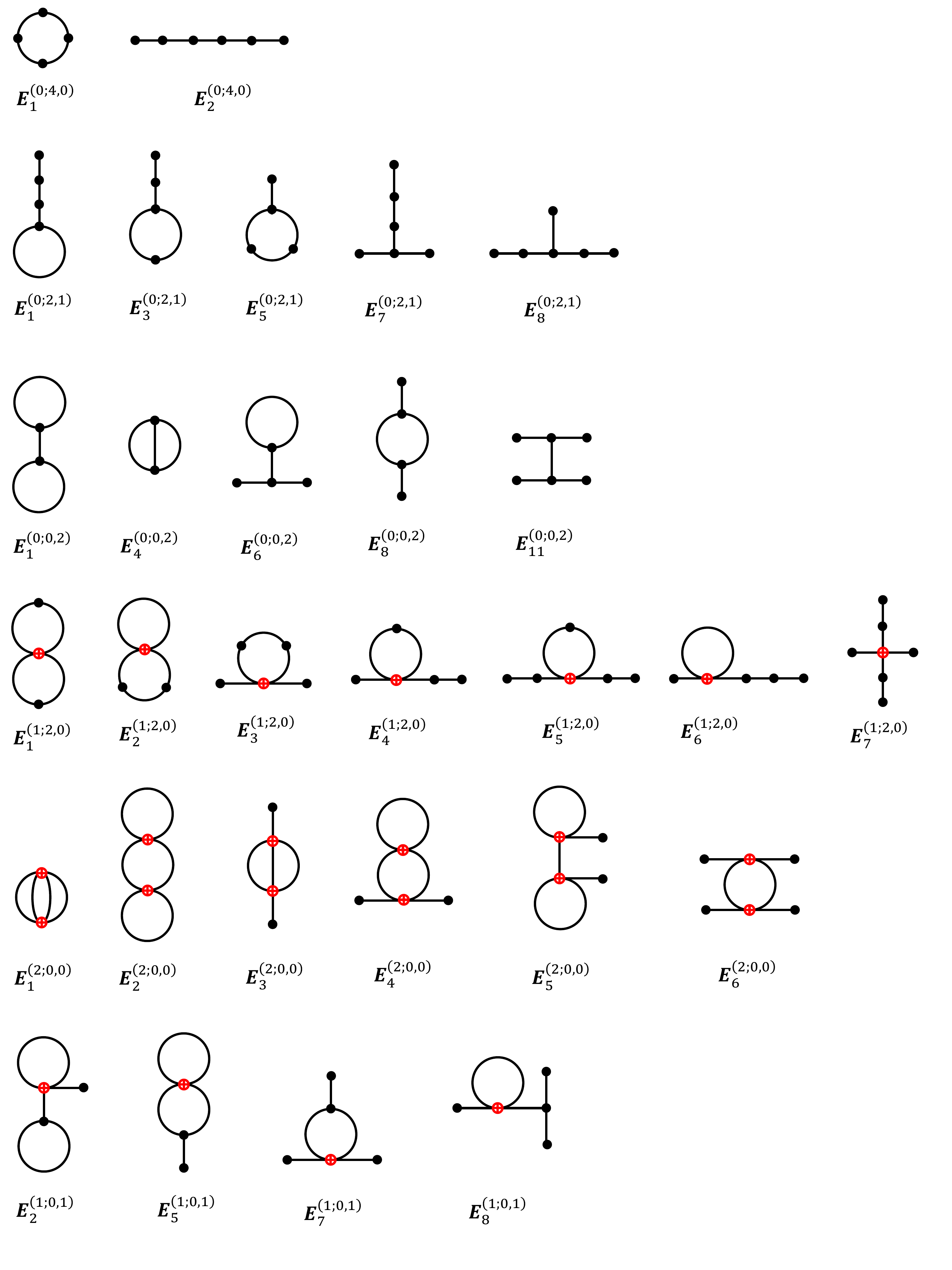}
		\par\end{center}
	\caption{Diagrammatic representation of the 29 unfactorizable parity preserving monomials for $d=4$.}
	\label{fig:d4}
\end{figure}

We are now ready to derive the complete basis for the monomials with $d=4$.
First we have to suppress those monomials that are not linearly independent after taking into account the antisymmetry of Levi-Civita tensor as well as the fact that Riemann tensor is the commutator of two covariant derivatives.
After some manipulations, we find the following linear dependences among various monomials
	\begin{eqnarray}
	\bm{E}_{2}^{\left(0;2,1\right)} & \equiv & \bm{E}_{1}^{\left(0;2,1\right)}+\bm{E}_{6}^{\left(1;2,0\right)},\\
	\bm{E}_{4}^{\left(0;2,1\right)} & \equiv & \bm{E}_{3}^{\left(0;2,1\right)}+\bm{E}_{4}^{\left(1;2,0\right)},\\
	\bm{E}_{6}^{\left(0;2,1\right)} & \equiv & \bm{E}_{5}^{\left(0;2,1\right)}+\bm{E}_{3}^{\left(1;2,0\right)},\\
	\bm{E}_{9}^{\left(0;2,1\right)} & \equiv & \bm{E}_{8}^{\left(0;2,1\right)}+\bm{E}_{7}^{\left(1;2,0\right)},
	\end{eqnarray}
	\begin{eqnarray}
	\bm{E}_{2}^{\left(0;0,2\right)} & \equiv & \bm{E}_{1}^{\left(0;0,2\right)}+2\,\bm{E}_{2}^{\left(1;0,1\right)}+\bm{E}_{5}^{\left(2;0,0\right)},\\
	\bm{E}_{3}^{\left(0;0,2\right)} & \equiv & \bm{E}_{1}^{\left(0;0,2\right)}+\bm{E}_{2}^{\left(1;0,1\right)},\\
	\bm{E}_{5}^{\left(0;0,2\right)} & \equiv & \bm{E}_{4}^{\left(0;0,2\right)}-\frac{1}{2}\bm{E}_{3}^{\left(2;0,0\right)},\\
	\bm{E}_{7}^{\left(0;0,2\right)} & \equiv & \bm{E}_{6}^{\left(0;0,2\right)}+\bm{E}_{8}^{\left(1;0,1\right)},\\
	\bm{E}_{9}^{\left(0;0,2\right)} & \equiv & \bm{E}_{8}^{\left(0;0,2\right)}+\bm{E}_{6}^{\left(2;0,0\right)}-2\,\bm{E}_{7}^{\left(1;0,1\right)},\\
	\bm{E}_{10}^{\left(0;0,2\right)} & \equiv & \bm{E}_{8}^{\left(0;0,2\right)}-\bm{E}_{7}^{\left(1;0,1\right)},
	\end{eqnarray}
and
	\begin{eqnarray}
	\bm{E}_{1}^{\left(1;0,1\right)} & \equiv & -\frac{1}{2}\bm{E}_{3}^{\left(2;0,0\right)},\\
	\bm{E}_{3}^{\left(1;0,1\right)} & \equiv & \bm{E}_{2}^{\left(1;0,1\right)}+\bm{E}_{5}^{\left(2;0,0\right)},\\
	\bm{E}_{4}^{\left(1;0,1\right)} & \equiv & \bm{E}_{5}^{\left(1;0,1\right)}-\bm{E}_{4}^{\left(2;0,0\right)},\\
	\bm{E}_{6}^{\left(1;0,1\right)} & \equiv & \bm{E}_{7}^{\left(1;0,1\right)}-\bm{E}_{6}^{\left(2;0,0\right)}.
	\end{eqnarray}
As a result, the complete basis for parity preserving monomials with $d=4$ consists of 29 unfactorizable monomials, which are
	\begin{eqnarray}
	&  & \bm{E}_{1}^{\left(0;4,0\right)},\qquad\bm{E}_{2}^{\left(0;4,0\right)}, \nonumber\\
	&  & \bm{E}_{1}^{\left(0;2,1\right)},\qquad\bm{E}_{3}^{\left(0;2,1\right)},\qquad\bm{E}_{5}^{\left(0;2,1\right)},\qquad\bm{E}_{7}^{\left(0;2,1\right)},\qquad\bm{E}_{8}^{\left(0;2,1\right)},\nonumber\\
	&  & \bm{E}_{1}^{\left(0;0,2\right)},\qquad\bm{E}_{4}^{\left(0;0,2\right)},\qquad\bm{E}_{6}^{\left(0;0,2\right)},\qquad\bm{E}_{8}^{\left(0;0,2\right)},\qquad\bm{E}_{11}^{\left(0;0,2\right)},\nonumber\\
	&  & \bm{E}_{1}^{\left(1;2,0\right)},\qquad\bm{E}_{2}^{\left(1;2,0\right)},\qquad\bm{E}_{3}^{\left(1;2,0\right)},\qquad\bm{E}_{4}^{\left(1;2,0\right)},\qquad\bm{E}_{5}^{\left(1;2,0\right)},\qquad\bm{E}_{6}^{\left(1;2,0\right)},\qquad\bm{E}_{7}^{\left(1;2,0\right)},\nonumber\\
	&  & \bm{E}_{1}^{\left(2;0,0\right)},\qquad\bm{E}_{2}^{\left(2;0,0\right)},\qquad\bm{E}_{3}^{\left(2;0,0\right)},\qquad\bm{E}_{4}^{\left(2;0,0\right)},\qquad\bm{E}_{5}^{\left(2;0,0\right)},\qquad\bm{E}_{6}^{\left(2;0,0\right)},\nonumber\\
	&  & \bm{E}_{2}^{\left(1;0,1\right)},\qquad\bm{E}_{5}^{\left(1;0,1\right)},\qquad\bm{E}_{7}^{\left(1;0,1\right)},\qquad\bm{E}_{8}^{\left(1;0,1\right)},  \label{cb_d4}
	\end{eqnarray}
together with $18+25+6+16+3+6=74$ factorizable monomials, of which the expressions can be read straightforwardly.
The diagrammatic representations of the 29 unfactorizable monomials are shown in Figure \ref{fig:d4}.

\begin{figure}[H]
		\begin{center}
			\includegraphics[scale=0.5]{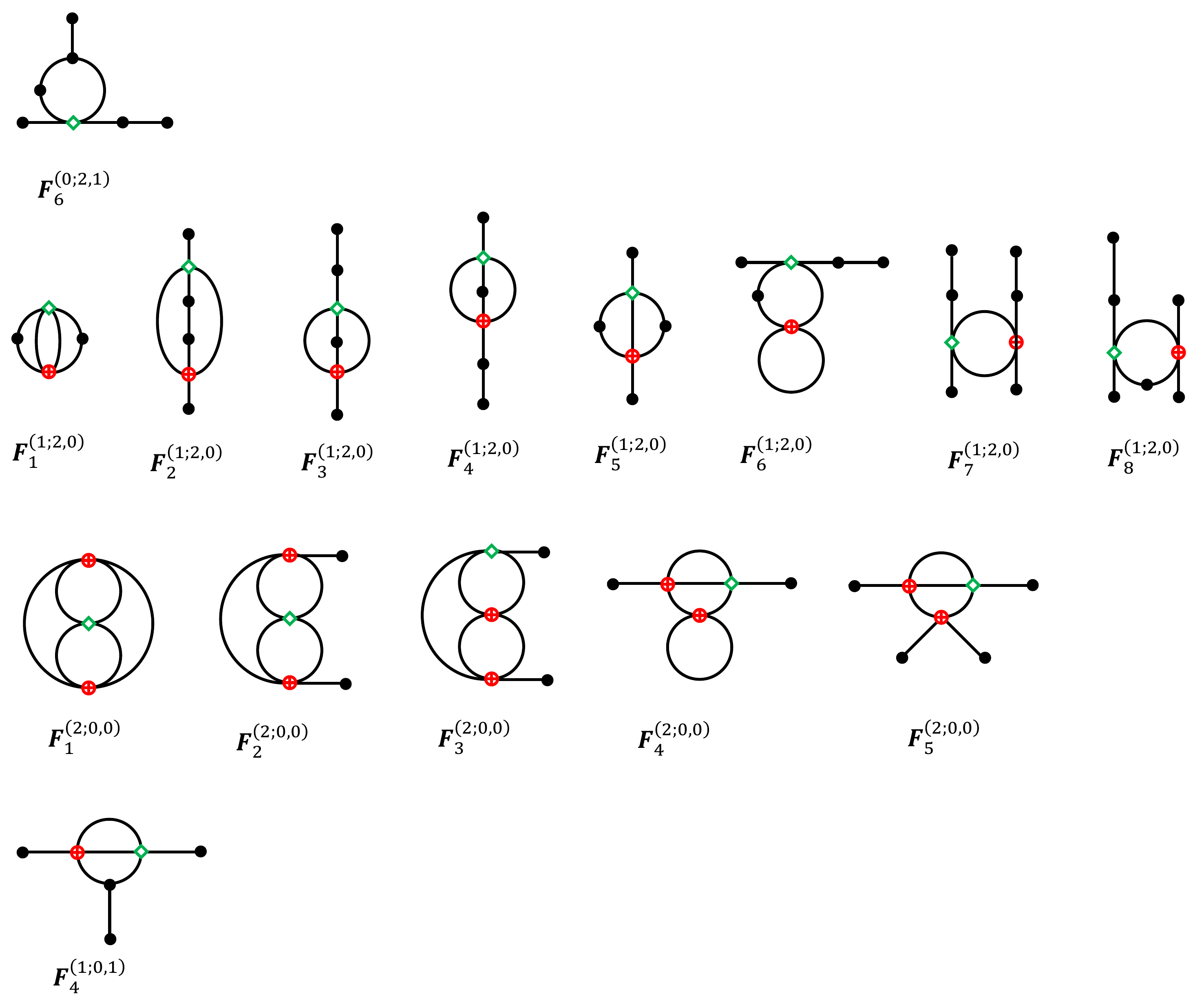}
			\par\end{center}
		\caption{Diagrammatic representation of the 15 unfactorizable parity violating monomials with $d=4$.}
		\label{fig:d4p}
	\end{figure}

In the case of parity violation, we have the following linear dependence among various monomials
	\begin{eqnarray}
	\bm{F}_{1}^{\left(0;2,1\right)} & \equiv & \bm{F}_{5}^{\left(1;2,0\right)},\\
	\bm{F}_{2}^{\left(0;2,1\right)} & \equiv & -\frac{1}{2}\bm{F}_{2}^{\left(1;2,0\right)},\\
	\bm{F}_{3}^{\left(0;2,1\right)} & \equiv & \frac{1}{2}\bm{F}_{3}^{\left(1;2,0\right)},\\
	\bm{F}_{4}^{\left(0;2,1\right)} & \equiv & -\frac{1}{2}\bm{F}_{7}^{\left(1;2,0\right)},\\
	\bm{F}_{5}^{\left(0;2,1\right)} & \equiv & \bm{F}_{6}^{\left(0;2,1\right)}-\bm{F}_{8}^{\left(1;2,0\right)},
	\end{eqnarray}
	\begin{eqnarray}
	\bm{F}_{1}^{\left(0;0,2\right)} & \equiv & -\frac{1}{4}\bm{F}_{2}^{\left(2;0,0\right)},\\
	\bm{F}_{2}^{\left(0;0,2\right)} & \equiv & -\frac{1}{2}\bm{F}_{4}^{\left(1;0,1\right)},\\
	\bm{F}_{3}^{\left(0;0,2\right)} & \equiv & -\frac{1}{2}\bm{F}_{4}^{\left(1;0,1\right)}+\frac{1}{2}\bm{F}_{5}^{\left(2;0,0\right)},
	\end{eqnarray}
and
	\begin{eqnarray}
	\bm{F}_{1}^{\left(1;0,1\right)} & \equiv & \frac{1}{2}\bm{F}_{2}^{\left(2;0,0\right)},\\
	\bm{F}_{2}^{\left(1;0,1\right)} & \equiv & -\frac{1}{2}\bm{F}_{3}^{\left(2;0,0\right)},\\
	\bm{F}_{3}^{\left(1;0,1\right)} & \equiv & \frac{1}{2}\bm{F}_{4}^{\left(2;0,0\right)},\\
	\bm{F}_{5}^{\left(1;0,1\right)} & \equiv & \bm{F}_{4}^{\left(1;0,1\right)}-\bm{F}_{5}^{\left(2;0,0\right)},\\
	\bm{F}_{6}^{\left(1;0,1\right)} & \equiv & \frac{1}{2}\bm{F}_{5}^{\left(2;0,0\right)}.
	\end{eqnarray}
Thus the complete basis for parity violating monomials with $d=4$ consists of 15 unfactorizable monomials
	\begin{eqnarray}
	&  & \bm{F}_{6}^{\left(0;2,1\right)},\nonumber\\
	&  & \bm{F}_{1}^{\left(1;2,0\right)},\qquad\bm{F}_{2}^{\left(1;2,0\right)},\qquad\bm{F}_{3}^{\left(1;2,0\right)},\qquad\bm{F}_{4}^{\left(1;2,0\right)},\qquad\bm{F}_{5}^{\left(1;2,0\right)},\qquad\bm{F}_{6}^{\left(1;2,0\right)},\qquad\bm{F}_{7}^{\left(1;2,0\right)},\qquad\bm{F}_{8}^{\left(1;2,0\right)},\nonumber\\
	&  & \bm{F}_{1}^{\left(2;0,0\right)},\qquad\bm{F}_{2}^{\left(2;0,0\right)},\qquad\bm{F}_{3}^{\left(2;0,0\right)},\qquad\bm{F}_{4}^{\left(2;0,0\right)},\qquad\bm{F}_{5}^{\left(2;0,0\right)},\nonumber\\
	&  & \bm{F}_{4}^{\left(1;0,1\right)}, \label{cb_d4p}
	\end{eqnarray}
together with $2+2=4$ factorizable monomials.
The diagrammatic representations of the 15 unfactorizable monomials are shown in Figure \ref{fig:d4p}.

\section{Conclusion} \label{sec:con}

In this work, we investigated the necessity and possibility of extending the scalar-tensor theory by including the third or even higher order derivatives of the scalar field as well as more general couplings among the curvature and higher derivatives of the scalar field.
Ghostfree higher curvature terms have been studied in the literature  \cite{Deruelle:2012xv,Crisostomi:2017ugk}.
From the point of view of the effective field theory, the third or even higher order covariant derivatives of the scalar field are of the same order as higher curvature terms.
Thus a full investigation of all possible monomials built of both higher order curvature terms and derivatives of the scalar field is necessary.

As being described in Sec. \ref{sec:form} in details, we assign each monomial with a set of integers $(c_0,c_1,\cdots;d_2,d_3,\cdots)$, which are the numbers of different orders of derivatives of the Riemann tensor and of the scalar fields, respectively.
The hierarchy of monomials is made according to the integer $d$ defined in (\ref{d_def}). 
For each $d$, we also classify monomials into different categories according to the integers $(c_0,c_1,\cdots;d_2,d_3,\cdots)$ of higher derivatives of the Riemann tensor and the scalar field. 
This classification is summarized in Table \ref{tab:clsST}, which is one of the main results in this work.
We argue that all monomials with the same value of $d$ are of the same order and thus should be treated in the same footing.
This not only explains the natural arising of derivatives of the scalar field beyond the second order, but also indicates that novel ghost-free Lagrangians may exist by combining higher curvature terms and higher derivatives of the scalar field \cite{Gao:toappear}.

In Sections \ref{sec:d1}-\ref{sec:d4}, we made a systematic and complete investigation of all the monomials for $d=1,2,3,4$.
We concentrated on the unfactorizable monomials in the irreducible cases, and derive their explicit expressions for each category $(c_0;d_2,d_3)$. 
Both parity preserving and parity violating cases are discussed.
The main results in this work are the complete basis for the monomials with $d=2,3,4$ present at the end of each section.
Due to the complexity and large amount of monomials, we also developed a diagrammatic representation of the monomials, which may help us to construct and classify terms in a transparent manner.
The diagrammatic representations of the unfactorizable monomials in the complete basis are present at the end of each section.

The results derived in this work will be the starting point of exploring more general viable higher derivative scalar-tensor theories, which we will present in the near future.

\acknowledgments

I would like to thank M. Crisostomi for correspondence.
This work was supported by the National Youth Thousand Talents Program of China and the Natural Science Foundation of China (NSFC) under the grant No. 11975020.

%

\end{document}